\documentclass[fleqn,usenatbib]{mnras}

\usepackage[T1]{fontenc}
\usepackage{ae,aecompl}
\usepackage[]{natbib}

\usepackage[pdftex]{graphicx}	
\usepackage{amsmath}	
\usepackage{amssymb}	
\usepackage{xfrac}
\usepackage{epstopdf}
\usepackage{float}

\title[Faint Satellites in the COSMOS Survey]{Quantifying the abundance of faint, low-redshift satellite galaxies in the COSMOS survey}

\author[Xi et al.]{
ChengYu Xi,$^{1}$\thanks{E-mail: cxi@uwaterloo.ca}
James E. Taylor,$^{1}$\thanks{E-mail: taylor@uwaterloo.ca}
Richard J. Massey,$^{2}$
\newauthor{
Jason Rhodes,$^{3,4}$
Anton Koekemoer,$^{5}$
and Mara Salvato$^{6,7}$}
\\
$^{1}$Department of Physics and Astronomy, University of Waterloo, 200 University Avenue West, Waterloo, Ontario, N2L 3G1, Canada\\
$^{2}$Institute for Computational Cosmology, Durham University, South Road, Durham, DH1 3LE, UK\\
$^{3}$California Institute of Technology, MC 249-17, 1200 East California Boulevard, Pasadena, CA 91125, USA\\
$^{4}$Jet Propulsion Laboratory, California Institute of Technology, Pasadena, CA 91109, USA\\
$^{5}$Space Telescope Science Institute 3700 San Martin Drive, Baltimore MD 21218, USA\\
$^{6}$MPE, Giessenbachstrasse 1, Garching 85748, Germany\\
$^{7}$Cluster of Excellence, Boltzmann Strasse 2, 85748, Germany}
\date{Accepted 2018 May 11. Received 2018 May 10; in original form 2017 December 29}

\pubyear{2018}

\begin{document}
\label{firstpage}
\pagerange{\pageref{firstpage}--\pageref{lastpage}}
\maketitle

\begin{abstract}
{    Faint dwarf satellite galaxies are important as tracers of small-scale structure, but remain poorly characterized outside the Local Group, due to the difficulty of identifying them consistently at larger distances. We review a recently proposed method for estimating the average satellite population around a given sample of nearby bright galaxies, using a combination of size and magnitude cuts (to select low-redshift dwarf galaxies preferentially) and clustering measurements (to estimate the fraction of true satellites in the cut sample). We test this method using the high-precision photometric redshift catalog of the COSMOS survey, exploring the effect of specific cuts on the clustering signal. The most effective of the size-magnitude cuts considered recover the clustering signal around low-redshift primaries (z < 0.15) with about two-thirds of the signal and 80\%\ of the signal-to-noise ratio obtainable using the full COSMOS photometric redshifts. These cuts are also fairly efficient, with more than one third of the selected objects being clustered satellites. We conclude that structural selection represents a useful tool in characterizing dwarf populations to fainter magnitudes and/or over larger areas than are feasible with spectroscopic surveys.} In reviewing the low-redshift content of the COSMOS field, we also note the existence of several dozen objects that appear resolved or partially resolved in the HST imaging, and are confirmed to be local (at distances of $\sim$250 Mpc or less) by their photometric or spectroscopic redshifts. This underlines the potential for future space-based surveys to reveal local populations of intrinsically faint galaxies through imaging alone.
\end{abstract}


\section{Introduction}
\label{sec:1}

The Milky Way, M31, and other bright galaxies in the nearby universe are observed to have retinues of faint dwarf satellites. The `classical' dwarfs of the Local Group, those identified decades ago, have magnitudes brighter than $M\sim-6$ in the $B$ or $V$-band, while the more recently discovered `ultra-faints' can be many magnitudes fainter \cite[see][for a review]{M12}. Given their high velocity dispersions and implied high mass-to-light ratios, dwarf satellites are inferred to trace the dense substructure seen in simulated dark matter halos. As such, they provide a very important test of models of structure formation. The relationship between dwarf satellites and halo substructure is complex, however, since the simplest models relating the two fail to match the number \citep{Klypin99, Moore99}, spatial distribution \citep{Kravtsov04} and central densities \citep{BoylanKolchin11} of the known dwarf galaxies of the Local Group. Detailed, careful modelling \citep[e.g.][]{Brooks14,Sawala15} seems to be required to understand the properties of these objects.

Despite ongoing observational efforts \citep[e.g.][and references therein]{K13, Chiboucas13, Sand14, Merritt14, Javanmardi16, Crno16, Mueller17, Greco2018}, most of our information about faint satellites comes from the Local Group, and models of dwarf galaxy formation typically set out to reproduce its properties. Our view of the Local Group is limited, however, by obscuration and uneven (albeit gradually improving -- \citealt{Laevens15a,Laevens15b,DES15a,DES15b}) sky coverage. Furthermore, studies of bright satellites around Milky Way analogues suggest that our Galaxy may be unusual in some respects; for instance, the presence of two bright, star forming satellites represents a 1 in 250 or rarer occurrence \citep{Robotham12}. To reach robust conclusions about typical satellite populations, we really need to expand the inventory of host systems with well-sampled satellite distributions by a factor of 100 or more.

Identifying faint satellites around more distant primaries is, unfortunately, very challenging. Over a reasonably large volume, such objects should be bright enough to be detected in large-area surveys. But without some means of determining distances to faint galaxies, and thus of associating them with nearby bright ones, local satellites are swamped by the much larger number of faint background galaxies. Recent work by the SAGA survey \citep{SAGA} provides a good indication of the challenge; a massive spectroscopic campaign measuring more than 17,000 redshifts found only two dozen new dwarfs down to a magnitude of $-12$, within a projected virial radius around their nearby targets. Going fainter would decrease the efficiency further, at prohibitive cost in observing  time. 

There are alternatives to spectroscopic distance determinations. Photometric redshifts are one example; in cases where many bands are available, these can be quite effective for determining 3D structure, at least on large scales \citep{Scoville_LSS,Scoville13}. Unfortunately, photometric redshifts of this quality are only available for a few small fields, notably the COSMOS field \citep{Scoville07}. A second possibility is to use clustering to estimate the average distance to a faint population of objects, by association with a brighter set of objects of known distance or redshift. Association inferred from proximity on the sky provides distance estimates for a number of the (relatively rare) dwarf galaxies in the `Local Volume' out to 11 Mpc \citep{K13}, while the related statistical technique of `clustering redshifts' \citep{Menard13, Rahman15} has been used to determine mean redshifts for populations at greater distances \citep[e.g.][]{Rahman16}. Even here, without any further sample selection beyond a basic magnitude cut, the clustering signal from faint, nearby systems tends to be weak.

A third alternative for estimating distances (or at least selecting local galaxies preferentially) is {    structural (size, magnitude, and/or surface-brightness)} selection. As we will show below, the intrinsically faint galaxies of the local universe occupy a distinct region of {    structural} parameter space. Cuts in {    size, magnitude,} and/or surface brightness are not enough to uniquely identify them at all redshifts, but can be quite effective for nearby objects. This method has been used implicitly several times, e.g.~in \cite{NGVS} or \cite{K13}, but without much systematic study. 
In \cite[][ST14 hereafter]{ST14}\defcitealias{ST14}{ST14}, we proposed a specific {    structural} selection criterion, based on size and magnitude, to identify satellites around primaries at distances of 10--40 Mpc. We demonstrated that our {    structural} cuts preferentially select nearby dwarfs by measuring the clustering signal of the cut sample with respect to the primaries. Overall, our cuts increased the signal-to-noise ratio (SNR) of the clustering signal from undetectable levels up to a value $\sim 9$, allowing us to measure a number of properties of the satellite population. One major limitation of this method, however, is the incompleteness of the resulting samples (down to a fixed magnitude or luminosity limit), which we estimated to be 50\%\  or more. Furthermore, our selection was tuned to relatively nearby systems. It is unclear how well this selection method extends to fainter magnitudes, and more generally, how it depends on the detailed form of the {    structural} cut.  

The goal of the current work is to study and test the {    structural} selection method in more detail. Ideally, we would do this with large ($\gtrsim 10^4$ object), complete samples of faint (21--22 magnitude) galaxies with measured spectroscopic redshifts. Unfortunately, no samples of sufficient depth and {    areal} coverage are currently available. 
The closest equivalent is the photometric redshift catalog of the COSMOS survey \citep{Mobasher07, Ilbert09, Laigle16}. This provides photo-$z$s with an 
accuracy of 1\%\ or better down to magnitudes of $i^{+}=23$ (or even deeper at low redshift), and thus gives a good indication of which faint galaxies are 
truly local, albeit over a very small field.
Since the COSMOS field is so small, we will push the limits of the selection method developed in \citetalias{ST14}, extending the distance range considered 
by a factor of 25, in order to increase the size of the primary sample and allow a robust detection of the clustering signal.

We apply {    structural} selection to galaxies in the COSMOS field, using various cuts based on {    structural} properties measured at ground-based 
resolution by the Sloan Digital Sky Survey \cite[SDSS hereafter --][]{SDSS00}{   , for consistency with \citetalias{ST14}}. Defining samples of bright, nearby primaries with spectroscopic redshifts 
and fainter secondary samples selected {    structurally},  we measure the clustering of secondaries with respect to primaries, and use this to estimate 
what fraction of the secondaries are true satellites. We study the effect of several different selection cuts
on the purity and completeness of the resulting satellites samples.

The outline of the paper is as follows. In Section 2, we first introduce the surveys and datasets used. In Section 3, we then present the basic argument behind {    structural} selection, using known dwarf populations from the Local Group or the `Local Volume' within 11 Mpc to estimate the intrinsic distribution of dwarf galaxy properties. In Section 4, we describe our selection of primary and secondary samples in the COSMOS field, and explain how the primary-secondary clustering amplitude is measured. To establish a baseline for the effectiveness of {    structural} selection, in Section 5, we measure this clustering amplitude for a secondary sample with no {    structural} cuts, as well
as for a sample with photo-$z$ cuts to isolate those objects most likely to lie at the distance of the primary. In Section 6, we apply cuts on secondary structure instead, 
and show how much of the clustering signal these can recover. Finally, in Section 7 we consider the very nearest systems in the COSMOS catalog, that appear to be resolved or partially resolved in the Hubble Space Telescope (HST) imaging available for the field \citep{Koekemoer2007}, and give an indication of the samples that future wide-field, space-based surveys 
will provide. In Section 8 we conclude by discussing the limitations of {    structural} selection, and the future prospects for this technique. Throughout the paper we calculate distances assuming a cosmological model with parameters $\Omega_{m,0} = 0.31$, $\Omega_{\Lambda,0} = 0.69$ and $h = 0.678$, consistent with recent Planck analyses \citep{Planck2013, Planck2015}.

\section{Data}
\label{sec:2}

The data considered in this paper includes several local samples of nearby galaxies, and the galaxies of the COSMOS field. For the latter, we use information both from high-resolution and/or space-based imaging, and from lower-resolution SDSS imaging. Each of the catalogs or sets of measurements is described below.

\subsection{The Local Group}

Although the inventory of identified local galaxies is always expanding, the Nearby Galaxy Catalog of \citet[][M12 hereafter]{M12}\defcitealias{M12}{M12} provides a reasonably recent summary of all known objects, up to a few Mpc from the Milky Way. We use the version of the catalog available on the author's web-site{\footnote{https://www.astrosci.ca/users/alan/Nearby\_Dwarfs\_Database.html}}, which was last updated in 2013. {    We have verified that with a few exceptions (e.g.~Canis Major), the objects in this version of the Nearby Galaxy Catalog also appear in the Local Volume Catalog described below. For internal consistency, we will use the distances, magnitudes and isophotal radii recorded in the latter, since it contains more objects overall. We will use the different size measurements given in the two catalogs to estimate a half-light radius for every object in the Local Volume Catalogue, as described below.}

\subsection{The Local Volume}

As discussed in the {    Introduction}, identifying distant dwarf galaxies is challenging, and current catalogs of nearby galaxies are probably very incomplete. The most extensive 
list of nearby systems beyond the Local Group is the `Local Volume Catalog' (LVC), first described in \citep{K04}. This catalogue was updated in (\citealt[][]{K13}\defcitealias{K13}{K13} -- K13 hereafter), and is available on-line\footnote{https://www.sao.ru/lv/lvgdb}. We will use the LVC as an indication of what more distant dwarfs might look like from a {    structural} point of view. In particular, we will use the $B_T$ magnitudes and $a_{26}$ sizes given in the on-line database, and documented in \citetalias{K13}. These are, respectively, total magnitudes in the Johnson $B-$band, from various sources listed in the database, and diameters of the isophotal radius corresponding to 26.5 mag/arcsec$^2$ in the $B$ band, estimated visually and calibrated using light profiles, as described in \citepalias{K13}. (\citetalias{K13} also notes that for objects with a central surface brightness equal to or fainter than 26.5 mag/arcsec$^2$, the isophotal diameter definition no longer makes sense; in these cases the values listed in the LVC correspond instead to the exponential scale radius.)  

For typical objects with exponential profiles, the radius $r_{26} = a_{26}/2$ should be roughly equal to the effective radius $r_{\rm eff}$ {    (or `half-light radius' $r_h$ in \citetalias{M12})}. In principle, we could assume a specific radial profile for each object and convert more carefully from $a_{26}$ to the effective radius, but we will not need this level of precision for the general arguments presented here. Comparing the LVC $r_{26}$ values to the effective radii $r_{\rm eff}$ for the same objects given in the Nearby Galaxies Catalog, we find that the median ratio of the two radii is 1.05, although with large scatter and a systematic dependence on morphological type. Thus, in what follows we will assume $r_{\rm eff} = r_{26} = a_{26}/2$ for the LVC objects. When needed, we will calculate exponential scale radii $r_{\rm exp}$ assuming an exponential profile, such that $r_{\rm eff} = 1.678 r_{\rm exp}$.
We will also use the mean surface density interior to the effective radius, calculated as 
\begin{eqnarray}
\left<\mu\right>_{\rm eff} &=& m_{1/2} + 2.5 \log(\pi r_{\rm eff}^2)\nonumber\\ 
&=& m_{\rm tot} + 1.995 + 5\log(r_{\rm eff})
\end{eqnarray}
where $m_{1/2}$ and $m_{\rm tot}$ are the magnitudes corresponding to half the luminosity and the total luminosity, respectively.

\subsection{More Distant Objects}

At distances $D > 11$ Mpc and out to a few tens of Megaparsecs, the Extragalactic Distance Database \citep{Tully09}{\footnote{http://edd.ifa.hawaii.edu}} 
provides a summary of many of the objects with known distances. We use the `Cosmicflows-3' sample from the database 
\citep[][T16 hereafter]{Tully16}\defcitealias{Tully16}{T16} as indicative of the state of knowledge about galaxy populations in the distance range 10--50 Mpc.
The database version of this catalog includes total $B$-band magnitudes, as well as distances estimated as described in \citetalias{Tully16}.

\subsection{COSMOS} 

To test our {    structural} selection methods, we need {    uniform} imaging for a large sample of faint galaxies 
with reasonably accurate (e.g.~$\sigma_D \lesssim 100$\,Mpc) distance estimates. Given the difficulty of obtaining spectra for faint objects, photometric redshifts (photo-$z$s) provide the only realistic solution. While photo-$z$s derived from shallow, optical photometry in five or fewer bands are of little use at low redshift \citep[e.g.][]{ST14, SAGA}, those derived from {    deeper imaging with} large numbers of narrow- and intermediate-band filters across the ultraviolet, optical and infrared range can achieve accuracies of 1\%\  or less \citep[e.g.][]{Ilbert09}. The largest sample of accurate photo-$z$s is from the COSMOS survey \citep{Scoville07}, a deep, multi-wavelength survey of a  2 deg$^2$ equatorial field.

COSMOS photo-$z$s were derived by template fitting, as described in \cite{Mobasher07} and \cite{Ilbert09}. More recently, they have been updated with the addition of new, deeper NIR and IR data from the UltraVISTA \citep{McCracken12} and SPLASH (Spitzer Large Area Survey with Hyper-Suprime-Cam\footnote{http://splash.caltech.edu}) projects \citep{Laigle16}. In this paper, we will use this updated catalog (`COSMOS 2015' hereafter)\footnote{\label{cs2015_online}ftp://ftp.iap.fr/pub/from\_users/hjmcc/COSMOS2015} for our analysis. The quality of the photo-$z$s has been verified by comparing to a large number (50,000 or more) of  spectroscopic redshifts available in the COSMOS field, notably from the zCOSMOS-bright sample \citep{Lilly07}.  In the redshift range $z=0$--1.2, photo-$z$s for objects of magnitude $i^{+}_{AB} \le 22.5$ have an r.m.s. scatter of $\sigma = 0.7$\%\ with respect to the spectroscopic redshifts, while the fraction of `catastrophic failures' with relative errors $|z_p - z_s|/(1+z_s) > 0.15$ is 0.51\% .

\begin{figure}
\begin{center}
\vspace{-15mm}
        \includegraphics[width=0.9\columnwidth]{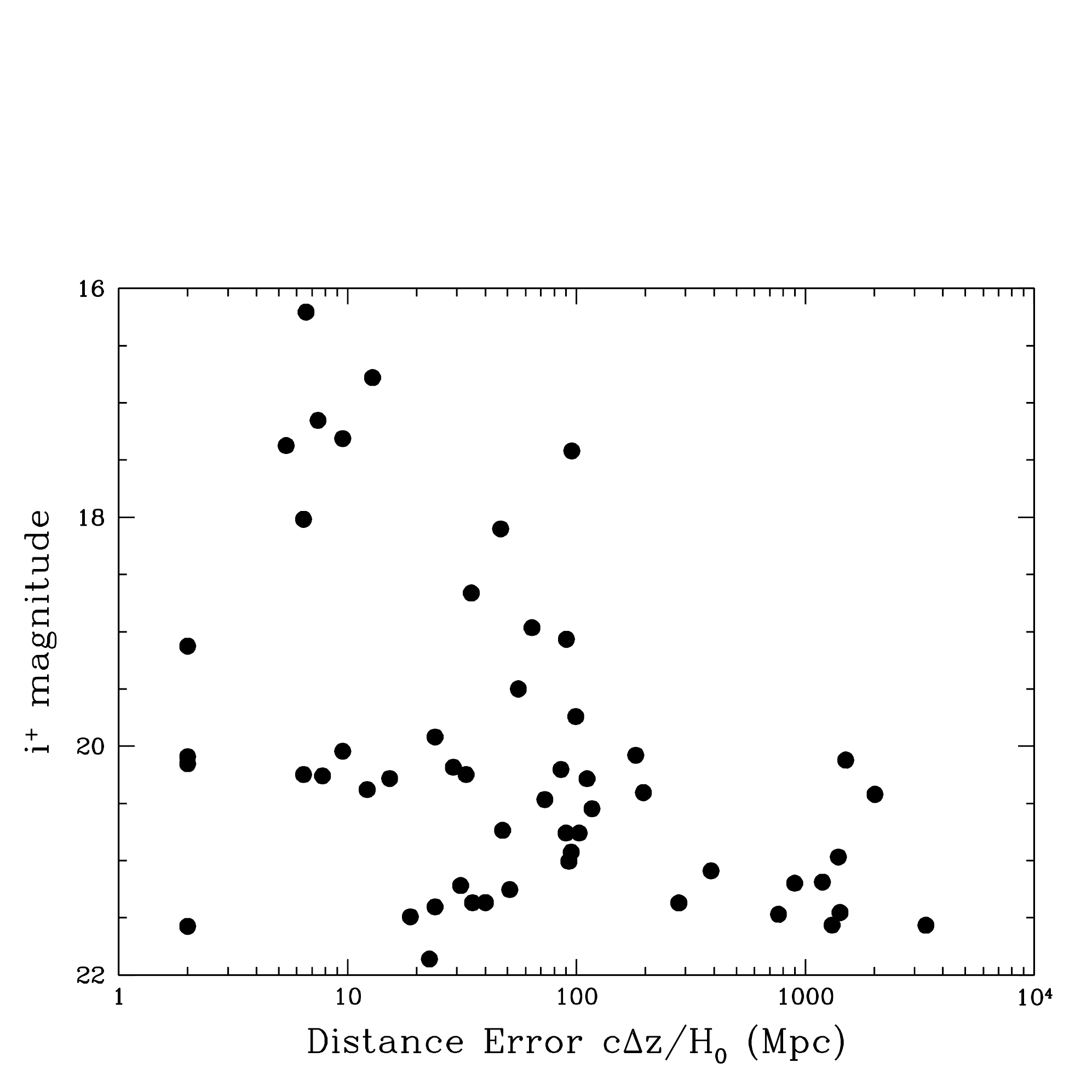}
    \caption{The absolute value of the difference between the photometric redshift and the spectroscopic redshift, converted to a distance error, for very local objects in the COSMOS 2015 catalog, as a function of their $i^{+}$ magnitude.}
    \label{fig:zerr}
\end{center}
\end{figure}

At very low redshift, these photometric redshift errors correspond to fairly small errors in distance. Figure~\ref{fig:zerr} shows the absolute value of the difference between the estimated photo-$z$ and the measured spectroscopic redshift, converted to a distance error using the approximation $\Delta D = c\Delta z/H_0$, for very local objects in the COSMOS 2015 catalog {    ($z_s < 0.06$)}, as a function of their $i^{+}$ magnitude. {    (Six objects have differences of zero to within roundoff errors, and have been placed at $\Delta D = 2$\,Mpc for clarity.)} We see that for most ($\sim$80\%) objects brighter than $i^{+} = 21$, the distance errors are less than 100 Mpc, and they are less than 40 Mpc for half the objects brighter than  $i^{+} = 22$. Thus, most bright objects from the catalog with very small photo-$z$s (e.g.~$z_p < 0.05$) should be genuinely nearby. We will return to this point in section~\ref{sec:7}.

In the process of fitting templates and estimating redshifts, \cite{Laigle16} also calculated stellar masses and star formation rates, which we will consider further below. Finally,  high-resolution imaging with ACS and/or WFC3 is available over most of the catalog area {    \citep{Koekemoer2007,CANDELS}}, via the IRSA cutout server\footnote{http://irsa.ipac.caltech.edu/data/COSMOS/index\_cutouts.html}. {    Convenient visual browsers for the ACS mosaic\footnote{https://www.mpia.de/COSMOS/skywalker} and the multi-wavelength coverage\footnote{http://www.cadc-ccda.hia-iha.nrc-cnrc.gc.ca/en/megapipe/cfhtls/scrollD2.html} of the field are also available}. 

While there is no single public redshift catalog for the COSMOS field, most measured redshifts for the field are now available via the NASA Extragalactic Database\footnote{https://ned.ipac.caltech.edu}. We have used these redshifts, and a few others available privately from the COSMOS collaboration, to correct the photo-$z$s when possible. Further work obtaining spectra in the COSMOS field is also ongoing, e.g.~with the C3R2 survey \citep{Masters2017}.

\subsection{SDSS Photometry in the COSMOS Field}
\label{subsec:SDSS}

The {    structural} selection initially introduced in \citetalias{ST14} was based on photometry from SDSS. SDSS covers a large area, but is both shallow (with a typical limiting magnitude of 22.2 in $r$\footnote{http://www.sdss.org/dr12/scope}) and has relatively poor seeing (a median value of 1.43${\arcsec}$ in $r$\footnote{http://classic.sdss.org/dr7/products/general/seeing.html}). Thus it represents an image quality easily achievable by other large, ground-based surveys. For a fair comparison with the results of \citetalias{ST14}, {    we will also use SDSS photometry here, querying} the spectroscopic and photometric galaxy catalogs from the latest SDSS Data Release 13 \citep[DR13 -- ][]{SDSS_DR13} and {    matching} the results to the COSMOS 2015 catalog.

To match catalogs, we first selected a subsample of COSMOS 2015 objects likely to have detections in SDSS. From the original catalogue of half a million objects, 
we selected objects with $i^{+} < 25.5$,  $0 < z < 6.9$, $\sigma_z < 0.5$, and $z - 2\sigma_z < 0.3$, where $i^{+}, z$ and $\sigma_z$ correspond to the catalogue quantities ${\tt IP\_MAG\_AUTO}$, ${\tt PHOTOZ}$, and ${\tt (ZPDF\_H68 - ZPDF\_L68)}/2$ respectively. These cuts produced a subsample of roughly 22,000 objects. 
We also queried the DR13 SkyServer\footnote{\label{sdss_sql}http://skyserver.sdss.org/dr13/en/tools/search/sql.aspx} to retrieve a photometric galaxy sample for the COSMOS region, with a `clean' cut (as described at \url{http://skyserver.sdss.org/dr13/en/tools/search/sql.aspx}) to ensure photometric quality. 
We then associated objects from the reduced COSMOS and SDSS catalogs with positions identical to within 1$\arcsec$ of each other. Galaxies without spectroscopic redshifts were assigned photometric redshifts and associated uncertainties from the COSMOS 2015 catalog, while those with spectroscopic redshifts were assigned the spectroscopic values, with an uncertainty of $\sigma_z = $0.0001. The resulting matched catalogue contains 12,108 objects.

For each object in the matched SDSS-COSMOS catalog, we obtained and saved ($r$-band) magnitudes and sizes from SDSS. To be consistent with \citetalias{ST14}, we used the composite model ({\tt cmodel}) magnitude, among the various magnitudes that SDSS provides. We did not apply a $K$-correction to these magnitudes, since our sample is relatively local. For galaxy sizes, we used the ($r$-band) exponential scale radius ({\tt expRad}) provided by SDSS, as in \citetalias{ST14}. Where necessary, we convert from this scale radius to an effective radius using the relation appropriate for an exponential profile, $r_{\rm eff} = 1.678 r_{\rm exp}$. The mean surface brightness within the effective radius is as calculated above, 
\begin{eqnarray}
\left<\mu\right>_{\rm eff} &=& m_{\rm tot} + 1.995 + 5\log(r_{\rm eff})\nonumber\\
{\rm or}\ \ \ \ \left<\mu\right>_{\rm eff} &=& m_{tot} +  3.1194 + 5\log(r_{\rm exp})\,,
\label{eq:reff}
\end{eqnarray}
once again assuming an exponential profile.

\section{The Basis for {    Structural} Selection}
\label{sec:3}

\begin{figure}
\includegraphics[width=\columnwidth]{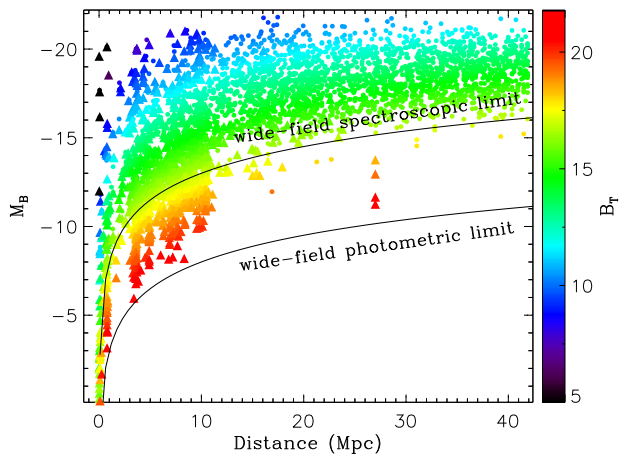}
\caption{Absolute $B$-band magnitude versus distance $D$ for nearby galaxies, from the catalogs of \citetalias{K13} and \citetalias{Tully16}. The colour scale shows the corresponding (total) apparent magnitude $B_{T}$. The upper and lower curves show the loci of objects with apparent magnitudes 17 and 22 respectively, roughly the limits for current, wide-field spectroscopic and photometric surveys such as SDSS.  Note the lack of objects between the spectroscopic and photometric limits with distances $D > 11$ Mpc. (Objects in a few known clusters in this distance range, such as Virgo and Fornax, are not shown on this plot.)}
\label{fig:1}
\end{figure}

Figure \ref{fig:1} shows a representative selection of nearby galaxies with distance estimates, including {LocalGroup/LVC and more distant objects from \citetalias{K13}} and \citetalias{Tully16} respectively. Galaxies are plotted in terms of their absolute $B$-band magnitude, estimated as described in Section 2. In general, both distances and magnitudes have considerable errors, particularly at faint magnitudes, but they give an indication of our knowledge of nearby galaxies. The upper and lower curves show the loci of objects with apparent magnitudes 17 and 22 respectively, roughly the completeness limits for current, wide-field spectroscopic and photometric surveys such as SDSS.

In the Local Group, at distances of less than {    3} Mpc, approximately 120 galaxies are known, including `ultrafaints' with absolute magnitudes $M_B > -6$. Within the Local Volume, the faintest identified objects generally correspond to the `classical' dwarfs of the Local Group, with magnitudes $ -15 < M_B < -6$. The total number of known objects in this volume is roughly 1000, though a comparison of the Local Group and LVC luminosity functions suggests the latter is incomplete by factor of up to 2 at $M_B = -10$, and a factor of 2--4 at the faintest magnitudes (see \citetalias{K13} for further discussion of the completeness of the LVC).
 
Beyond this there is a much larger volume, out to distances of 40--50 Mpc, where classical dwarfs should be easily detectable in the photometric catalogs of large-area surveys such as SDSS, given their photometric limits $M\sim 22$ (lower curve), but will lie below the typical spectroscopic limits of these surveys ($M\sim 17$ -- upper curve). Given the number of objects identified in the Local Volume, for instance, we might expect $\sim$ (4--5)$^3$ times as many, or $\sim $ 100,000 galaxies, most of them dwarfs, out to $D = 50$ Mpc. On the other hand, at faint magnitudes background counts will overwhelm these local objects. The SAGA survey \citep{SAGA}, for instance, counts roughly 3000 galaxies  per square degree down to an extinction-corrected magnitude limit of $r_0 = 20.75$, versus the handful of nearby dwarf galaxies expected per square degree. Their spectroscopic follow-up around nearby bright galaxies obtained more than 17,000 spectra, but yielded only 25 new satellites, that is a detection rate of less than 1/500. This inefficiency raises the question of whether intrinsically faint, nearby galaxies could be preferentially selected by their photometric properties alone, and if so, over what range of distances.

\begin{figure}
\begin{center}
\includegraphics[width=1.0\columnwidth]{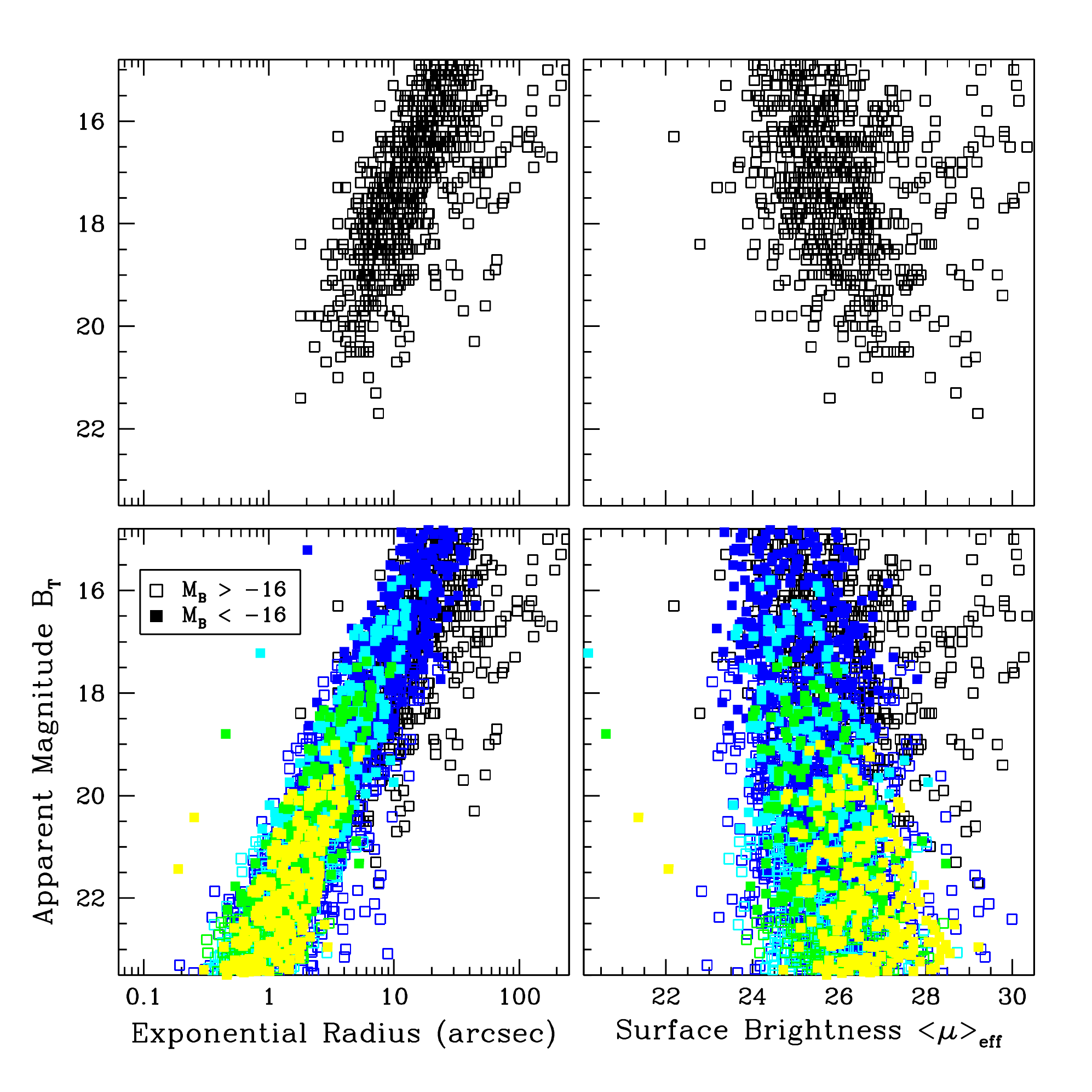}
\end{center}
\caption{{\it {    Bottom} Left Panel:} Apparent $B$-band magnitude versus apparent size, for LVC galaxies as seen at their original distances (D = 0--11 Mpc; black points), 
or at redshifts of 0.01--0.02 (blue), 0.05 (cyan), 0.1 (green), and 0.2--0.3 (yellow). Solid symbols indicate intrinsically bright galaxies ($M_B < -16$), while open symbols indicate intrinsically faint galaxies ($M_B > -16$). {\it {    Bottom} Right Panel:} Apparent  magnitude versus mean surface brightness $\left<\mu\right>_{\rm eff}$, for LVC galaxies seen at various distances. Symbols and colours are as in the left-hand panel. {    Top panels show the distributions of LVC galaxies at their original distances (i.e.~the black points) only, for clarity.} Note some regions of parameter space in either panel are dominated by low-redshift dwarf galaxies (open squares).}
\label{fig:3}
\end{figure}

One possible, albeit crude, alternative to complete spectroscopic surveys is to use the {    structural} properties of dwarfs to separate them from background galaxies. Galaxies in the nearby universe show a clear trend in surface brightness with intrinsic luminosity. At fixed apparent magnitude, intrinsically faint galaxies have lower mean surface brightness on average, or equivalently, larger angular sizes on the sky. Thus it may be possible to select them preferentially using size or surface-brightness cuts.
We can illustrate this by considering how the photometric properties of objects in the LVC sample of \citetalias{K13} would change if we saw them 
at progressively larger distances. The left and right panels of Figure~\ref{fig:3} show how apparent magnitude, size and surface brightness change as we move 
the LV sample from their original distances (D = 0--11 Mpc; black points) to redshifts of 0.01--0.02 (blue), 0.05 (cyan), 0.1 (green), or 0.2--0.3 (yellow). 
Solid squares indicate intrinsically bright galaxies ($M_B < -16$), while open squares indicate intrinsically faint galaxies ($M_B \ge -16$).

In the left hand panel, objects get smaller and fainter, moving down and to the left, as their redshift increases. Intrinsically faint galaxies (open symbols) are more diffuse than intrinsically bright ones, however; as a result, at fixed apparent magnitude, dwarfs are typically 2--3$\times$ larger than intrinsically bright galaxies on the sky. A cut in size 
and/or magnitude that selects the tail of the apparent size distribution will thus enhance the fraction of local, intrinsically faint galaxies in a sample.

The right hand panel shows a similar effect in magnitude versus surface brightness. With increasing redshift, objects move to fainter magnitudes, and then eventually shift to lower surface brightness as cosmological dimming becomes important. Intrinsically faint galaxies start at lower surface brightness, however, so the upper right hand side of 
the plot is dominated by low-redshift dwarf galaxies. Once again, cuts in surface brightness and/or magnitude may select these objects preferentially out of larger samples.

We note several caveats. First, the points in Figure~\ref{fig:3} show the locus of typical galaxies at each distance, but do not account for changing abundance due to the increasing volume probed at larger distances. Second, we have assumed that the sample of \citetalias{K13} is representative of cosmological volumes 
in general, while in fact some galaxy types (e.g.~those found in clusters) may be rare or missing entirely from the LVC sample. Finally, the region dominated 
by local dwarf galaxies in the right-hand panel lies at fairly low mean surface brightness. SDSS catalogs start to become significantly incomplete at central surface 
brightnesses of $\mu_0 \simeq$ 24--24.5 \citep{Blanton05}, although some objects can be recovered down to $\mu_0 \simeq$ 26--26.5 \citep{Kniazev04}. For the 
exponential surface-brightness profile typical of dwarf galaxies, these correspond to $\left<\mu\right>_{\rm eff} = $ 25.1--25.6 or $ \left<\mu\right>_{\rm eff }= $ 27.1--27.6 respectively. Thus, objects in the most interesting region of parameter space may not be detectable in conventional, shallow surveys such as SDSS. 

These complications motivate an empirical test of {    structural} selection, using the COSMOS photometric redshift catalog, one of the only samples with accurate distance estimates for large numbers (tens or hundreds of thousands) of faint galaxies. In what follows, we will apply various {    structural} cuts to this catalog and estimate their effect on the satellite population by measuring the resulting clustering signal.
 
\section{Clustering Measurement Method}
\label{sec:4}

To confirm that our {    structural} selection method works, we can measure the clustering of {    structural}ly-selected samples with respect to nearby bright galaxies that have spectroscopic redshifts, and thus reliable distance estimates.  A positive clustering signal will indicate that at least part of the {    structural}ly-selected sample lies at the same distance as the primary sample, and thus that we are preferentially selecting intrinsically faint, local galaxies. We describe the construction of the primary and secondary samples, the clustering measurement, and the corrections for masking below.

\subsection{Selecting Primaries} \label{sec:p_select}

Our goal in constructing the primary sample is to select bright galaxies similar to the Milky Way, at distances small enough that their satellites will be included in the COSMOS catalog, yet extending to high enough redshift that we have enough primaries to measure the clustering of their satellites with a reasonable SNR. We take as a starting point the photometry and photometric redshifts of the COSMOS 2015 photo-$z$ catalog \citep{Laigle16}, and proceed as follows:

\begin{enumerate}
\item First we select all galaxies with $M_{K_S} < -21.5$.
\item We then select those galaxies with photometric redshifts $z - 2\sigma_z < 0.3$, such that they have a reasonable chance of being low redshift objects 
(we choose a generous upper limit of $z=0.3$ at this stage to make sure we do not exclude any primaries at the upper end of our highest redshift range.)
\item For this subsample, we then check for any available spectroscopy, and correct the redshift if necessary, adjusting the absolute magnitudes correspondingly. 
For objects with spectroscopic redshifts, the redshift errors are assumed to be {$\sigma_z$ = 0.0001, or $\sigma_v$ = 30 km\,s$^{-1}$}.
\item Finally, we select only those galaxies with redshift errors $\sigma_z \le 0.1$ (removing two galaxies with redshifts $z > 0.3$ from the primary sample.)
\end{enumerate}
This selection process produces an initial sample of 735 primary galaxies. We estimate halo masses and virial radii for these objects from their stellar masses, assuming a standard stellar-to-halo mass relation \citep[e.g.][]{Leauthaud12}. {    The median stellar mass in this initial sample is $\langle M_*\rangle\sim2.5 \times 10^{10} M_\odot$, corresponding to a halo of mass $M_{h} \sim 10^{12} M_\odot$, with a virial radius $R_{200c}$\footnote{Where $R_{200c}$ is the radius within which the halo has a mean density 200 times the critical density $\rho_c$.} $\sim 200$\,kpc.} We find that a few of the nearest and most massive systems {    (with $M_* = 2$--$3 \times 10^{11} M_\odot$)} are predicted to have very large {    halo masses and} projected virial radii {    $R_{200c} > 1$ Mpc}, complicating the clustering calculations. Thus, we make an additional cut, removing from the sample objects with absolute magnitudes brighter than $-21.5$ in the (SDSS) $r$-band. This cut reduces the final number of primaries to 527{   , and the median stellar mass to $\langle M_*\rangle\sim2 \times 10^{10} M_\odot$. The largest stellar masses in the final cut sample are $M_*\sim 7 \times 10^{10} M_\odot$, and have estimated virial radii $R_{200c} < 300$\,kpc, such that our clustering calculations extend to more than three projected virial radii, even in the largest systems.}

The primary sample is then divided into three redshift ranges: 
\begin{itemize}
\item z=0.07--0.15, which contains 34 primaries;
\item z=0.15--0.20, which contains 57 primaries;
\item z=0.20--0.25, which contains 149 primaries.
\end{itemize}
The remaining 287 primaries have redshifts of 0.25 or more, which we will show is beyond the useful range for {    structural} selection.
The full redshift distribution of the primary sample is shown in Figure~\ref{fig:Np_Ns_vs_z_cumulative}. 

Finally, we note that \citetalias{ST14} also applied isolation cuts to their parent sample, to select primaries in the field or in poor groups (and thus close analogues of the Milky Way), as opposed to members of rich groups or clusters. Applying similar isolation cuts to our sample reduces the number of primaries considerably, so we will forego these cuts in the current paper, since the focus here is on testing structure as a distance indicator, rather than on characterizing the satellite population of a given type of primary. 

\subsection{Selecting Secondaries}	\label{sec:s_select}

\begin{figure}
\includegraphics[width=\columnwidth]{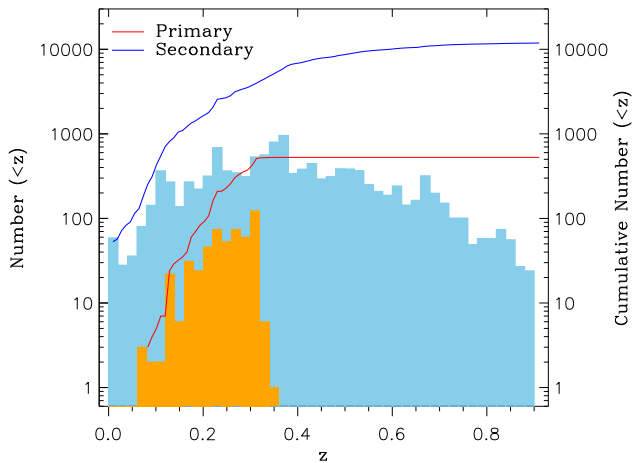}
\caption{The cumulative and differential redshift distributions of the primary (lower curve \& histogram) and secondary (upper curve and histogram) samples, over the range $z=0$--1. Note that a few secondaries lie beyond the redshift range shown on the plot.}
\label{fig:Np_Ns_vs_z_cumulative}
\end{figure}

Our secondary source catalogue consists of those objects we were able to match between the COSMOS 2015 and SDSS catalogues, as described in 
Section~\ref{subsec:SDSS}. This sample contains 12,108 objects in total. We do not place any further cuts on this sample, since our initial goal is to 
test how much of the clustering signal we can recover without additional information. The photometric redshift distribution of the secondary sample is shown in 
Figure~\ref{fig:Np_Ns_vs_z_cumulative}, over the range $z = 0$--1. (Note there are a few secondaries with redshifts beyond $z=1$ not shown on the plot.)

\subsection{Masking Corrections}

The COSMOS field includes regions with poor photometry in one or more bands, due to contaminating halos from bright stars, ghosts from internal reflection, or other artefacts. While detailed mask files for these regions exist in each of the 30+ COSMOS bands, we have found it less computationally demanding to calculate clustering using a single, approximate mask image with coarser sampling. We construct this mask empirically by making a 390 $\times$ 390 map of source counts in the COSMOS field, based on the entire (uncut) COSMOS 2015 catalog. Cells in this map with one or no counts are treated as potential masked regions. In a second round, any potential masked cell is determined to be masked if it has multiple neighbours with no counts. The resolution of our map file ($\sim 14\arcsec$) and threshold of one count were set such that the probability of masking a cell by chance due to Poisson fluctuations is extremely small (0.0026\%). We have experimented with variants on this method, changing the source count map resolution from 200 $\times$ 200 up to 600 $\times$ 600, and varying the threshold for counting cells as masked. We find that our clustering signals are stable to within $\sim5$\%\ with respect to these variations, but that the final mask looks most accurate for resolutions around the value (390 $\times$ 390) adopted here.

\subsection{The Clustering Calculation}

To measure clustering, we calculate the (cross-)correlation function of secondaries with respect to primaries, that is 
$$\xi(R_p) \equiv {{\Delta N}\over{N_{\rm exp}}}(R_p) = {{N_{\rm obs} - N_{\rm exp}}\over{N_{\rm exp}}}(R_p)$$
where $N_{\rm obs}$ is the number of primary-secondary pairs observed at separation $R_p$, $N_{\rm exp}$ is the number of pairs expected for a uniform distribution, and $R_p$ is the projected physical separation, assuming both members of the pair lie at the (spectroscopically determined) distance of the primary. We will also consider the `excess number', which is simply $\Delta N (R_p) = \xi N_{\rm exp}$.

Our method is essentially the same as that described in \citetalias{ST14}, with a few modifications in order to apply it at larger distances, so that we can obtain reasonable 
statistics given the relatively small field. First, we calculate the projected separations $R_p$ of all the primary-secondary pairs, assuming the secondaries lie at the same distance as the primary. We then count the number of pairs as a function of separation, in linear bins of width 50 kpc, ranging from 50 to 1000 kpc (with the bins centered on separations of 75 kpc, 125 kpc, etc.). 
The innermost bin ($R_p$ = 0--50 kpc, corresponding to 0--8.5$\arcsec$ at $z=0.25$) is excluded to avoid potential contamination from components (e.g.~\mbox{H\,{\sc ii}} regions) of the primary detected independently in the catalog, and because it is comparable to the resolution of our mask for the highest redshift primaries. 

To calculate the expected number of pairs, we use a local background density determined from the secondary counts between projected separations of $R_p =$\,600 and 950 kpc (this range is also consistent with \citetalias{ST14}). Given the stellar masses of our primaries, this range of separations should {    correspond to roughly 2--3 times the} virial radius of their halos, and therefore measures the larger-scale local background (the `2-halo term'), rather than the overdensity associated with the primary halo. The expected counts in the outer region are corrected for masking, and then scaled to the masked area of each inner bin to determine the expected number in that annulus.
  
The excess counts in bin $i$ around primary $j$ are thus:
\begin{equation}
\Delta N_{i,j} = \frac{A_{0,i}}{A_{i,j}} \left( N_{i,j} - \frac{A_{i,j}}{A_{{\rm{outer}},j}} N_{{\rm{outer}},j} \right)
\end{equation}
 where $A_{0,i} = 2\pi (R^2_{i} - R^2_{i-1})$ is the full geometric area of bin $i$ in the absence of masking (and assuming small angles),  $A_{i,j}$ is the area of bin $i$ around primary $j$ after masking,  $N_{i,j}$ are the total counts in bin $i$ around primary $j$,  $A_{{\rm{outer}},j}$ is the net area of the outer region used to calculate the background, after masking, and $N_{{\rm{outer}},j}$ are the total counts in this region. 
 
  \subsection{Figure of Merit for Clustering}
 \label{subsec:FOM}
 
 To quantify the extent to which {    structural} cuts can preferentially select local samples, it is convenient to define a single measurement of clustering that we can use as a figure of merit. In what follows, we will consider the SNR of the mean excess counts per primary $\Delta N$ (the `clustering signal'), summed over the range of separations $R_p = 50$--450 kpc, relative to a local background estimated from the secondary counts at separations $R_p = 600$--950 kpc. To calculate the error in $\Delta N$, we assume the main uncertainty in the mean excess counts comes from the Poisson errors on the galaxy counts $N^b_{\rm{inner}}$ and $N_{\rm{outer}}$, which are propagated into an error in the final value $\Delta N$ in the usual way. 
 
\section{The Clustering Signal}
\label{sec:5}

\subsection{The Signal with no Additional Cuts on the Secondary Sample}
\label{subsec:nc}

\begin{figure*}
	\includegraphics[width=1.9\columnwidth]{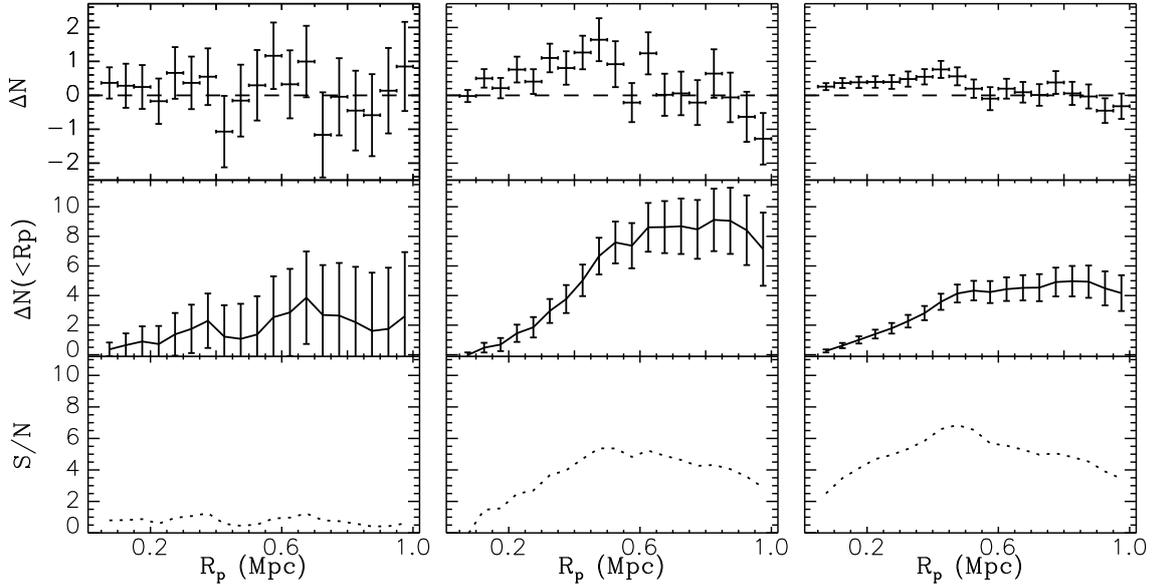}
	\caption{Clustering between the full secondary sample and primaries in the redshift ranges $z=0.07$--0.15, $z=0.15$--0.20, and $z=0.20$--0.25 (left, middle, and right plots respectively). In each plot, the three panels are, from top to bottom, the mean excess number of secondaries per primary in each radial bin, the cumulative excess number per primary as a function of radius (excluding objects with $R_p < 50$ kpc), and the total SNR of the cumulative excess detection.}
	\label{fig:ns_z015_z020_z025}
\end{figure*}

To establish a baseline for subsequent measurements, we first calculate the clustering signal $\Delta N$, by the method described in the previous section, using the entire secondary sample. Figure~\ref{fig:ns_z015_z020_z025} shows the clustering signal of the full secondary sample with respect to primaries in the three redshift ranges, $z=0.07$--$0.15$, $z=0.15$--$0.20$ and $z=0.20$--$0.25$ (left, middle, and right plots respectively). In each plot, the top panel shows the mean excess counts per primary in each annular bin; the middle panel shows the cumulative counts within $R_p$ (excluding objects at $R_p < 50$ kpc), and the bottom panel shows the SNR of the cumulative excess, given the uncertainties in the excess counts in individual bins. {    (Note that since $\Delta N$ can be negative in any given bin, the cumulative counts and SNR do not necessarily increase monotonically with radius.)}

In the lowest redshift range, we see that while there is some marginal evidence of clustering -- the differential counts interior to 600 kpc are positive on average -- the SNR of the cumulative excess is around 1 or less. We infer that more distance information is needed to determine which secondaries are associated with these nearby primaries, and to remove background galaxies from the secondary sample. The middle and upper redshift bins show stronger clustering, the SNR peaking at a value of 5.5--7, at projected separations $R_p = 450$--500 kpc. This scale corresponds to $\sim 1.5$ times the virial radius of our primaries, and matches the extent of the clustering signal seen in \citetalias{ST14}. In terms of our previously defined figure of merit, the SNR for $\Delta N$ cumulated over the range 50-450 kpc is 0.6, 4.5, and 6.4 for the three redshift ranges respectively.

\subsection{The Signal with Photo-$z$ Cuts on the Secondary Sample}
\label{subsec:photoz}

\begin{figure}
    \includegraphics[width=\columnwidth]{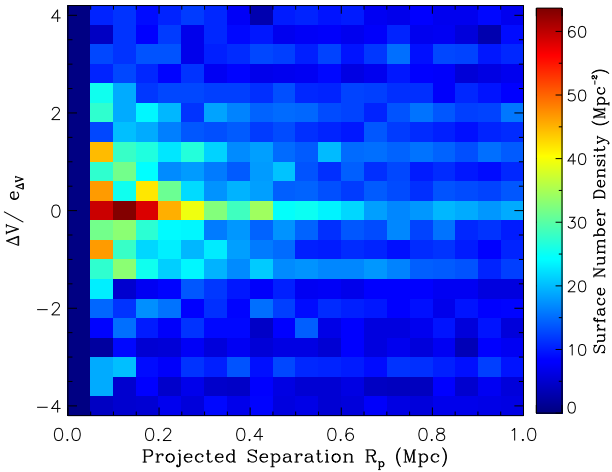}
    \caption{The surface number density of primary-secondary pairs as a function of projected separation $R_p$ and velocity offset $\Delta V = c\Delta z$, where the latter has been calculated from the photo-$z$s, and is expressed in units of the velocity uncertainty $e_{\Delta V} = c\sigma_z$.}
    \label{fig:sd_rp_dvedv}
\end{figure}

Whereas photometric redshifts derived from a few broad bands are of limited use at low redshift \citep[e.g.][]{SAGA},  the COSMOS photo-$z$s claim percent-level accuracies, even for relatively faint galaxies at low redshift. In Figure~\ref{fig:sd_rp_dvedv}, we test this accuracy. The plot shows the surface density of secondaries around primaries, as a function of projected separation $R_p$ and of velocity separation $\Delta V = c\Delta z$ as inferred from the photo-$z$s, the latter in units of the velocity/redshift error $e_{\Delta V} = c\sigma_z$ claimed in the catalog. We see a clear clustering signal at small projected separations, that is generally confined to the $\pm2\sigma_z$ range around the primary velocity. Assuming this excess corresponds to physically associated satellites, the width of the velocity offset distribution indicates that the photo-$z$ error estimates in the catalog are generally realistic.

\begin{figure*}
	\includegraphics[width=1.9\columnwidth]{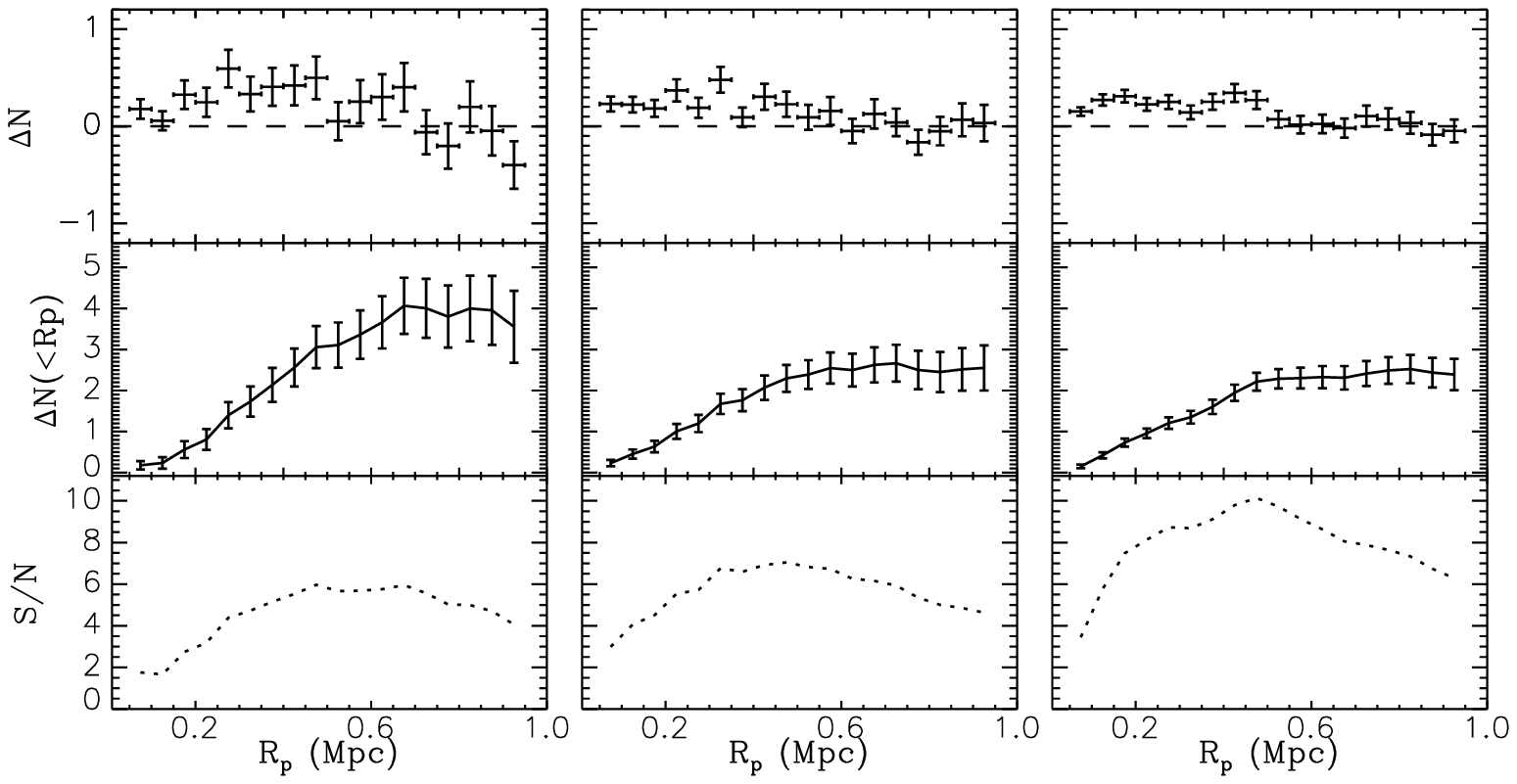}
	\caption{The clustering signal, as in Figure~\ref{fig:ns_z015_z020_z025}, but after applying photo-$z$ cuts to select only those secondaries
	likely to be at the same redshift as their primary.}
	\label{fig:ns_zcut_z015_z020_z025}
\end{figure*}

Given the validity of the photo-$z$ error estimates, we can select around each primary only those secondaries whose redshifts lie within $\pm 2\sigma_z$. (We note that secondaries should have real, physical velocity offsets with respect to the primary, but these will be negligible compared to the photo-$z$ errors, which are typically several thousand km\,s$^{-1}$.) The resulting clustering signal for this cut sample is shown in Figure \ref{fig:ns_zcut_z015_z020_z025}. Comparing Figures~\ref{fig:ns_z015_z020_z025} and \ref{fig:ns_zcut_z015_z020_z025}, we see that the photometric redshift cuts significantly improve the detection of the clustering signal, increasing the SNRs from less than 1 to 5.5, from 4.5 to 6.9, and from 6.4 to 9.8, in the three redshift ranges respectively. 

If the photo-$z$ selected sample were complete, these results and the results from Section~\ref{subsec:nc} would bracket the range 
of clustering amplitude and SNR we could expect from {    structural} selection. If photo-z selection is relatively inefficient, we will measure clustering around 
the different primary samples with SNRs comparable to those in the lower panels of Figure ~\ref{fig:ns_z015_z020_z025}, while if it is extremely 
efficient, we may approach the SNRs shown in the lower panels of Figure~\ref{fig:ns_zcut_z015_z020_z025}. (If the photo-$z$ selection is incomplete, e.g.~because 
of missing photo-$z$s or large redshift errors for certain objects, {    structural} selection could actually produce a larger amplitude signal than photo-z selection, albeit with lower a SNR.)

\section{Effect of {    Structural} Cuts}
\label{sec:6}

As shown in the previous section, the SNR of the clustering signal $\Delta N$ (the figure of merit defined in section~\ref{subsec:FOM}) can be increased significantly by removing 
background galaxies from the secondary sample. We test the effect of five simple, single-parameter {    structural} cuts, and two slightly more complicated 
two-parameter cuts, on the SNR of this measurement. 

\subsection{Single-parameter Cuts}
\label{subsec:6.1}

\begin{figure*}
	\includegraphics[width=0.6\columnwidth]{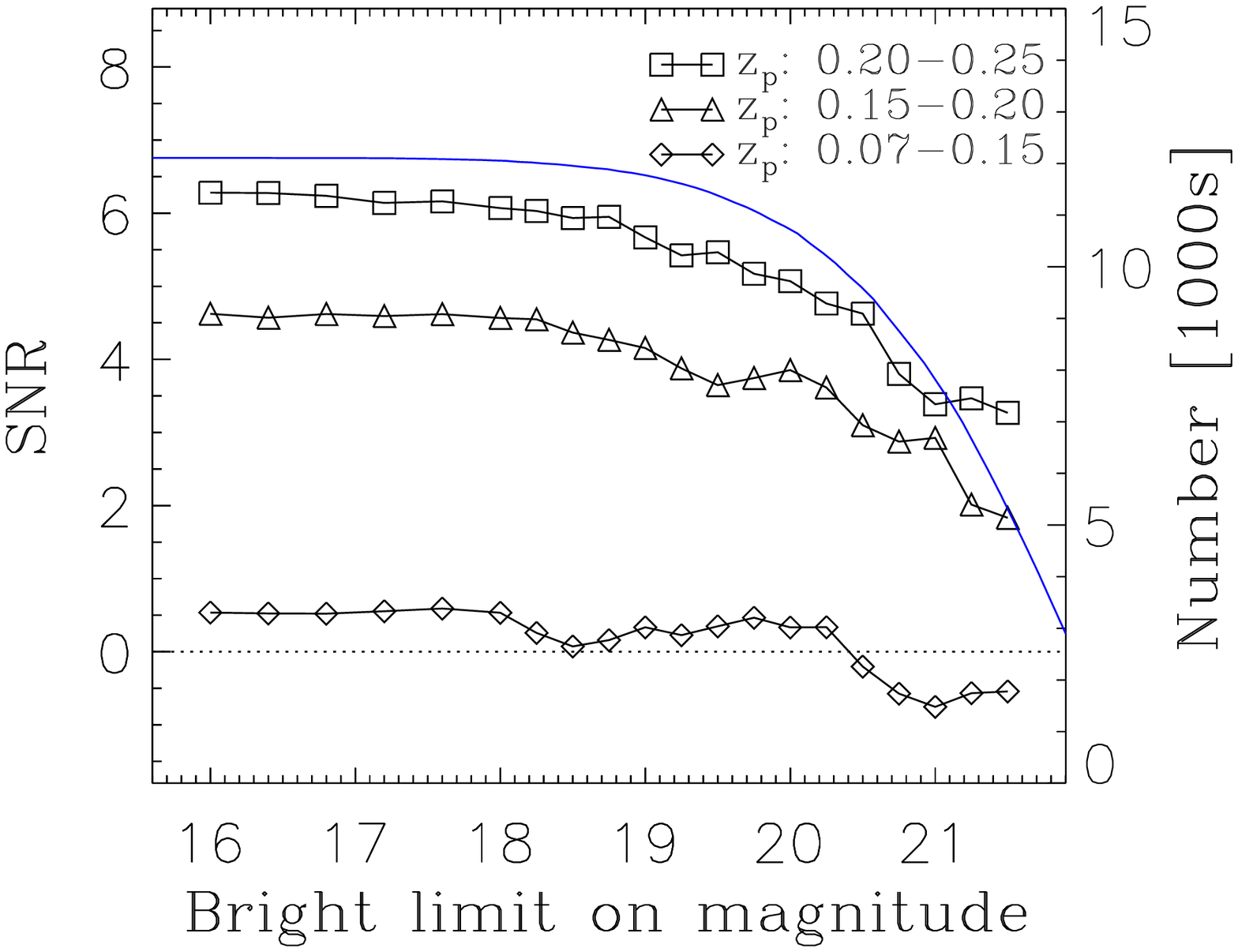}
	\includegraphics[width=0.6\columnwidth]{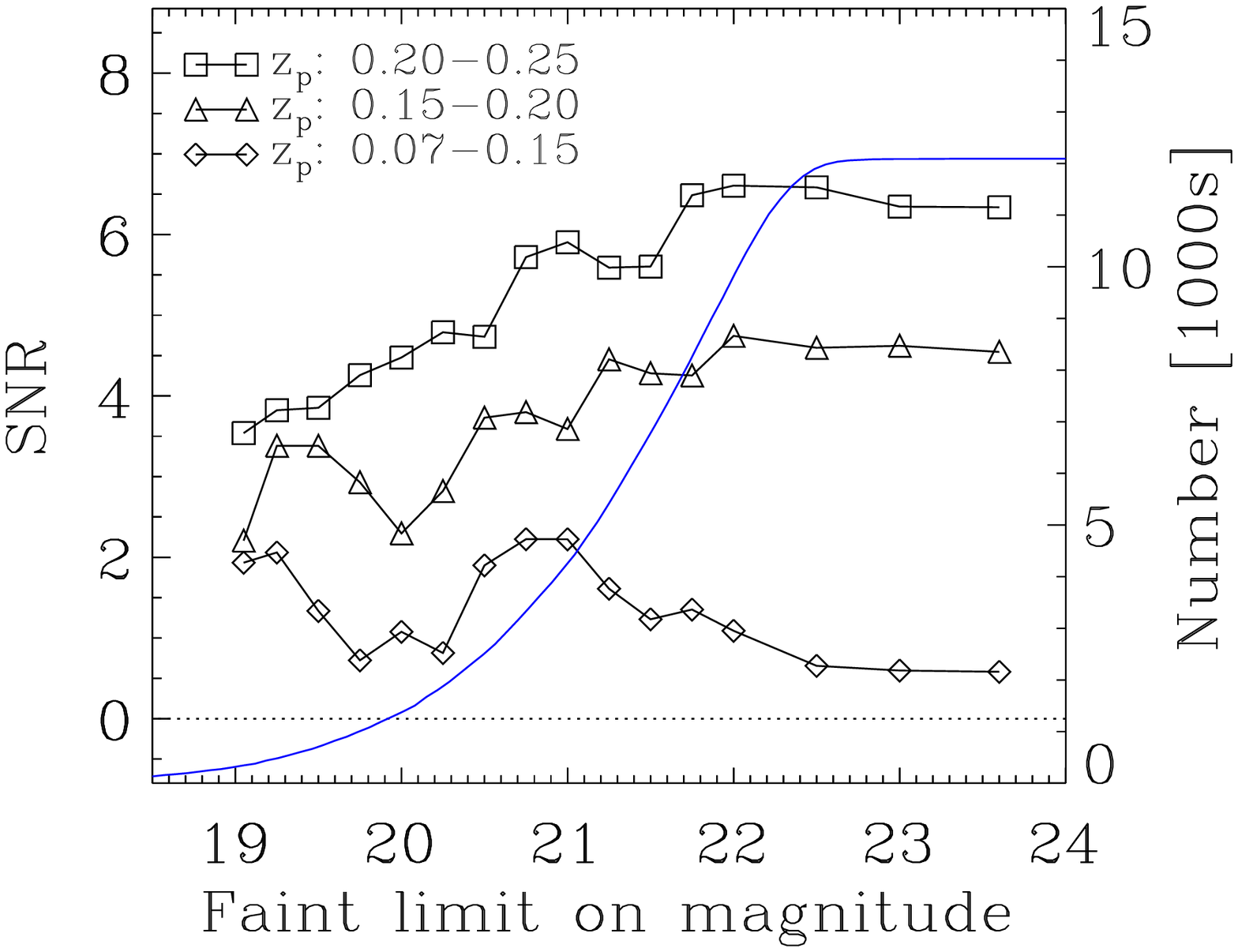}
	\includegraphics[width=0.6\columnwidth]{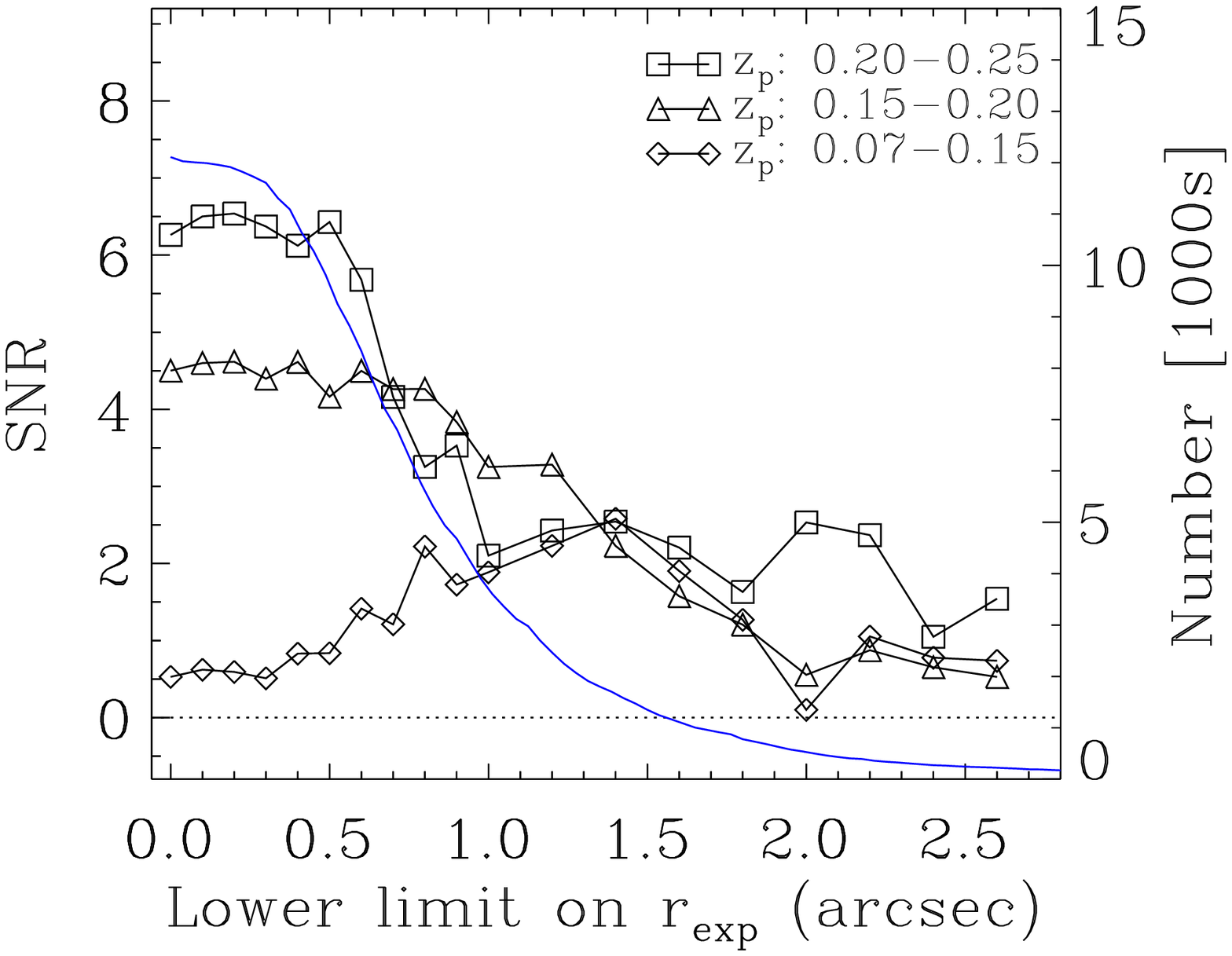}
	
	\includegraphics[width=0.6\columnwidth]{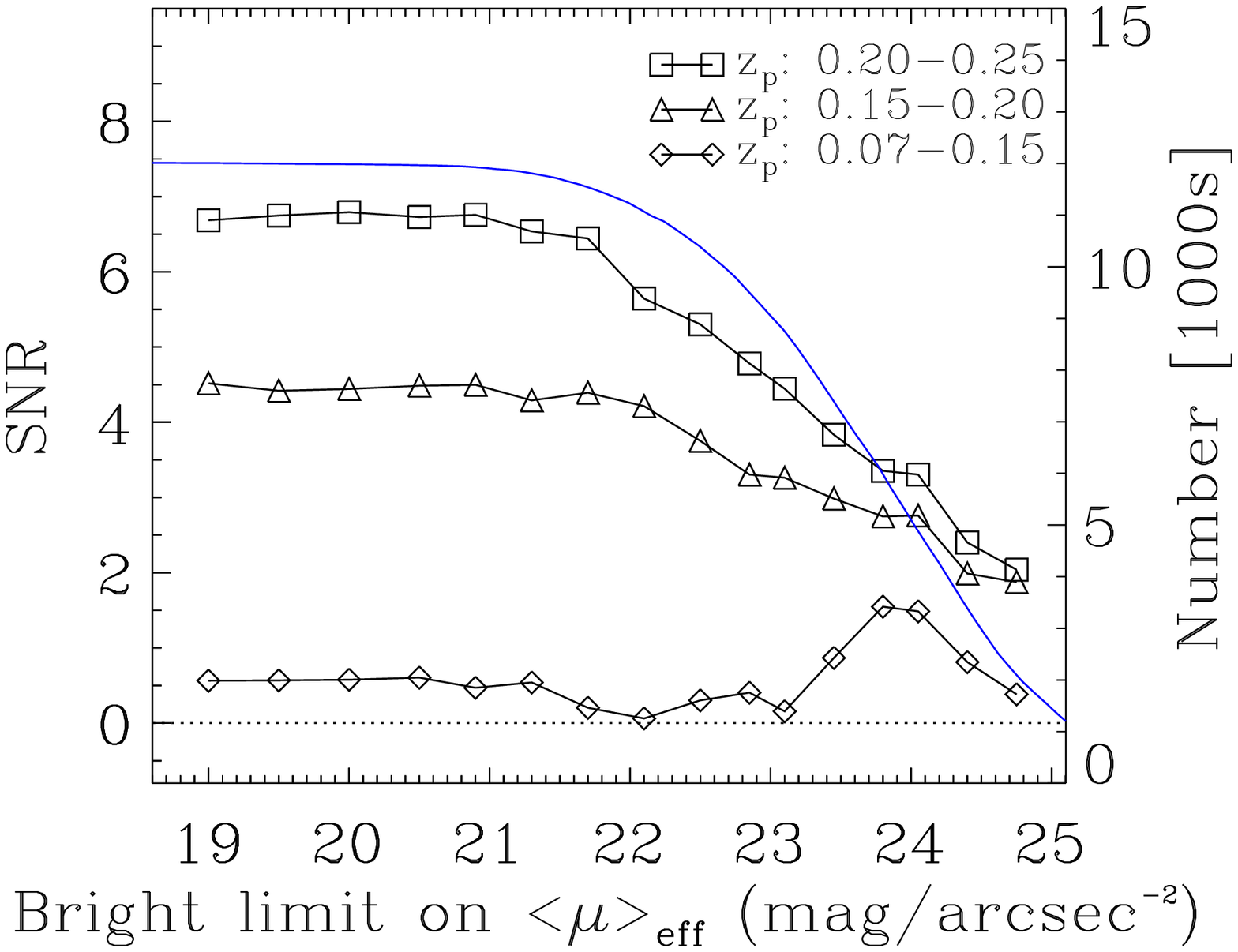}
	\includegraphics[width=0.6\columnwidth]{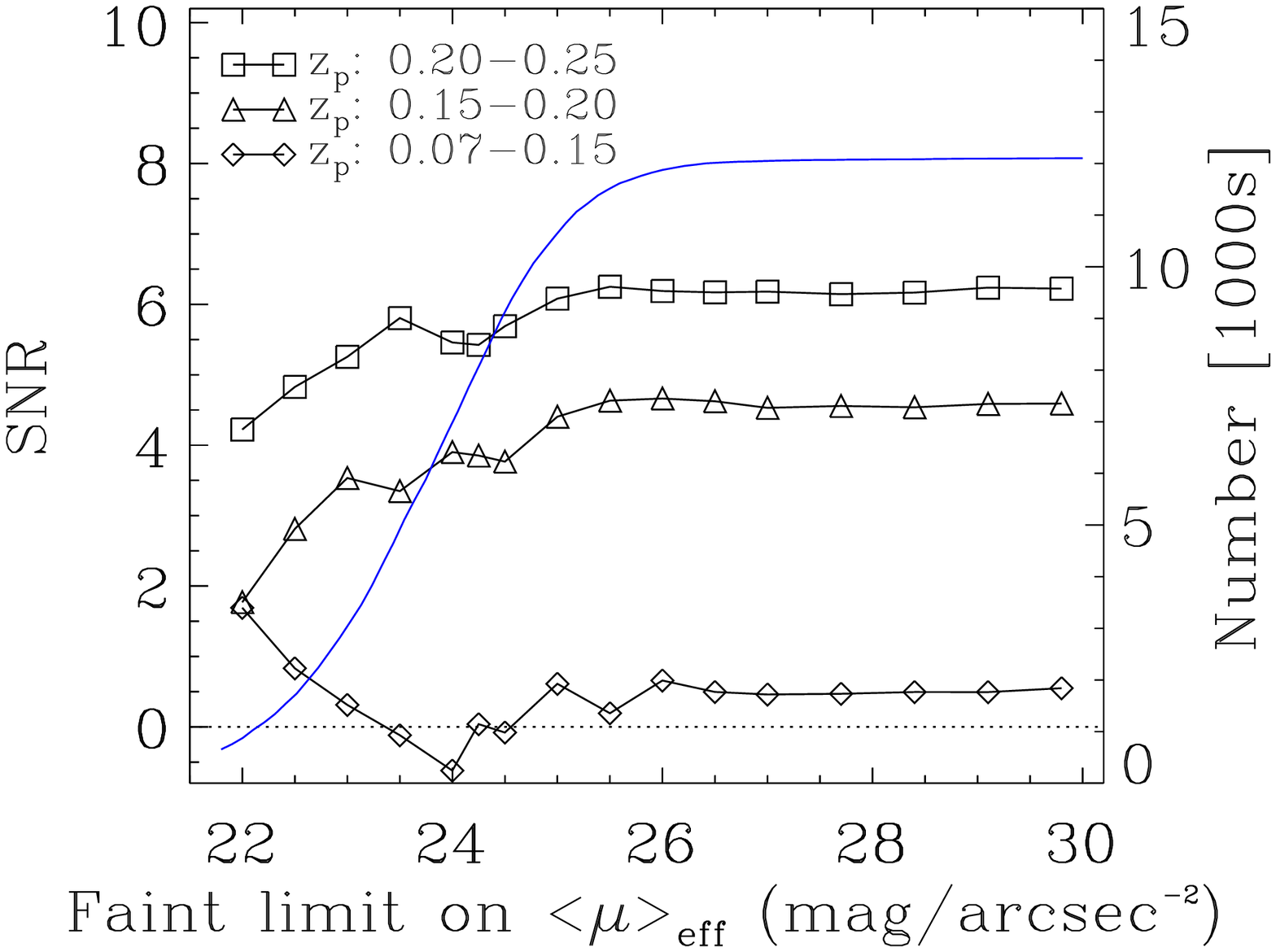}
	\caption{{    The clustering SNR as a function of bright and faint limits on the magnitude of the secondaries
		(top row, left and centre panels), a lower limit on secondary size $r_{\rm exp}$ (top row, right panel), 
		and  bright and faint limits on secondary surface brightness $\langle\mu\rangle_{\rm eff}$ 
		(bottom row, left and right panels). 
		All quantities are measured in the SDSS $r$-band.} 
		In each panel, the three sets of points are for the three primary redshift ranges. 
		The smooth (blue) curve in each panel indicates the total number of secondaries left 
		in the sample after applying the magnitude cut (with the scale indicated on the right side of the plot).}
	\label{fig:snr_vs_all}
\end{figure*}

The single-parameter cuts we test are:
\begin{itemize}
\item a cut on bright magnitudes, $r>r_{\rm{bright}}$

\item a cut on faint magnitudes, $r < r_{\rm{faint}}$

\item a cut on small sizes $r_{\rm exp} > r_{\rm exp}^{\rm{low}}$

\item a cut on high surface brightness, $\mu > \mu_{\rm{bright}}$

\item a cut on low surface brightness, $\mu < \mu_{\rm{faint}}$
\end{itemize}
These are shown in the five panels of Figure~\ref{fig:snr_vs_all}, from top left to bottom right.
In the latter two cases, the surface brightness is the mean value within the effective radius, $\left<\mu\right>_{\rm eff}$, as defined in Equation~\ref{eq:reff}.

Reviewing the results of the first two cuts, in the {    top left and middle panels} of Figure~\ref{fig:snr_vs_all}, we conclude that a bright magnitude limit on the secondary sample has little effect on the SNR, until this limit becomes faint enough that it starts reducing the size of the sample substantially (at which point the SNR drops correspondingly). A faint magnitude limit has more complicated effects. For the lowest redshift primaries, the maximum SNR is achieved by cutting out secondaries fainter than $r\sim21$, while for the higher redshift primary samples, a faint magnitude cut has little effect, provided it is fainter than $r\sim21$--21.5. We note however that as we make the magnitude cut fainter, the SNR increases before the size of the secondary sample does. We conclude that  objects brighter than $r\sim 21$ provide a large part of the clustering signal. (All of these results are of course contingent on the magnitude distribution of our secondary sample, which extends only to $\sim$22.5, since our secondaries are required to have SDSS photometry.)

The results of imposing a lower size limit depend on the redshift range of the primary sample {    (top right panel)}. At low redshifts ($z < 0.15$), the SNR starts out at $\sim0.5$, and increases to 2.5 as the value of the lower size limit increases to $1.5\arcsec$. The initial increase in SNR makes it clear that large objects are more often local, and that a lower size cut $r_{\rm exp} \gtrsim 1$--1.5$\arcsec$ can enhance the fraction of local dwarfs in the sample. If the size limit is increased beyond this value, the SNR drops, probably due to the loss of objects from the secondary sample (as indicated by the smooth curve). Also the effectiveness of size cuts is restricted to low redshift; for the two upper redshift bins, a lower limit on secondary size reduces the SNR of the clustering signal overall.

Somewhat surprisingly, limits on surface brightness $\mu$ do not generally improve the SNR, except possibly at low redshift. In the lowest redshift bin, cutting out objects with surface brightness $\mu \lesssim 24$ increases the clustering SNR from $\sim$0.5 to {    1.5}, but for the higher redshift bins, the highest SNR are achieved for no cuts at all.
(A faint cut around $\mu\sim22$ also appears to increase the SNR of the clustering measurement for the lowest redshift primaries, but in this case the size of the secondary sample is so small that we take this to be noise in the calculated SNR.) 
 
Overall, we conclude that for low-redshift primaries, a lower limit on secondary size and/or a faint limit on magnitude can significantly increase the SNR of the clustering measurement. At higher redshift ($z = 0.2$--0.25), single parameter cuts generally have no effect, or reduce the SNR. Given the distribution of local dwarfs in magnitude-surface brightness or magnitude-size space (Figure~\ref{fig:3}), we expect that {    simultaneous} cuts in two parameters may be more effective than single-parameter cuts. Before we consider these, however, we will briefly discuss the purity of the cut samples.

\subsection{Purity of the Secondary Samples}
 
\begin{figure*}
 	\includegraphics[width=0.9\columnwidth]{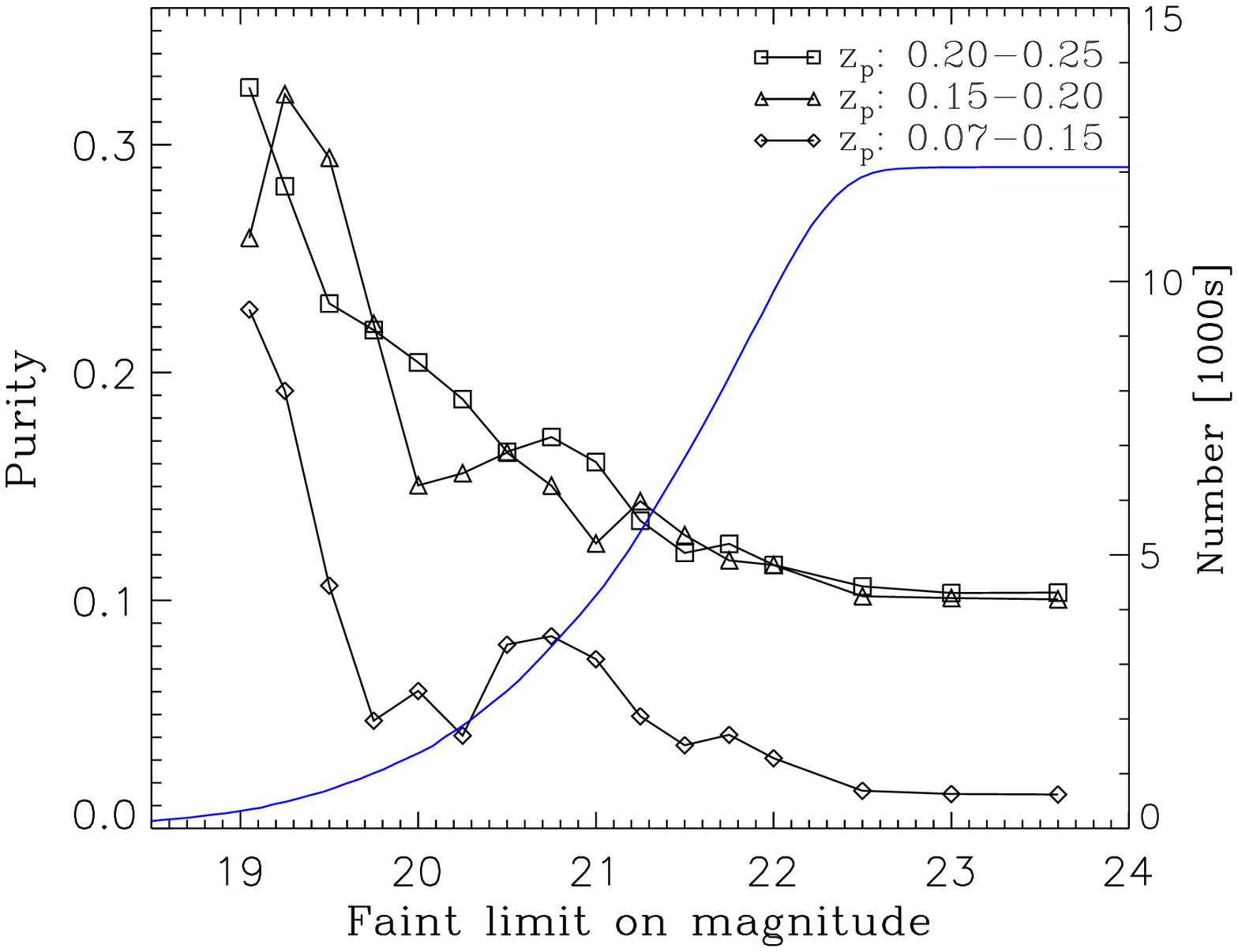}
	 \includegraphics[width=0.9\columnwidth]{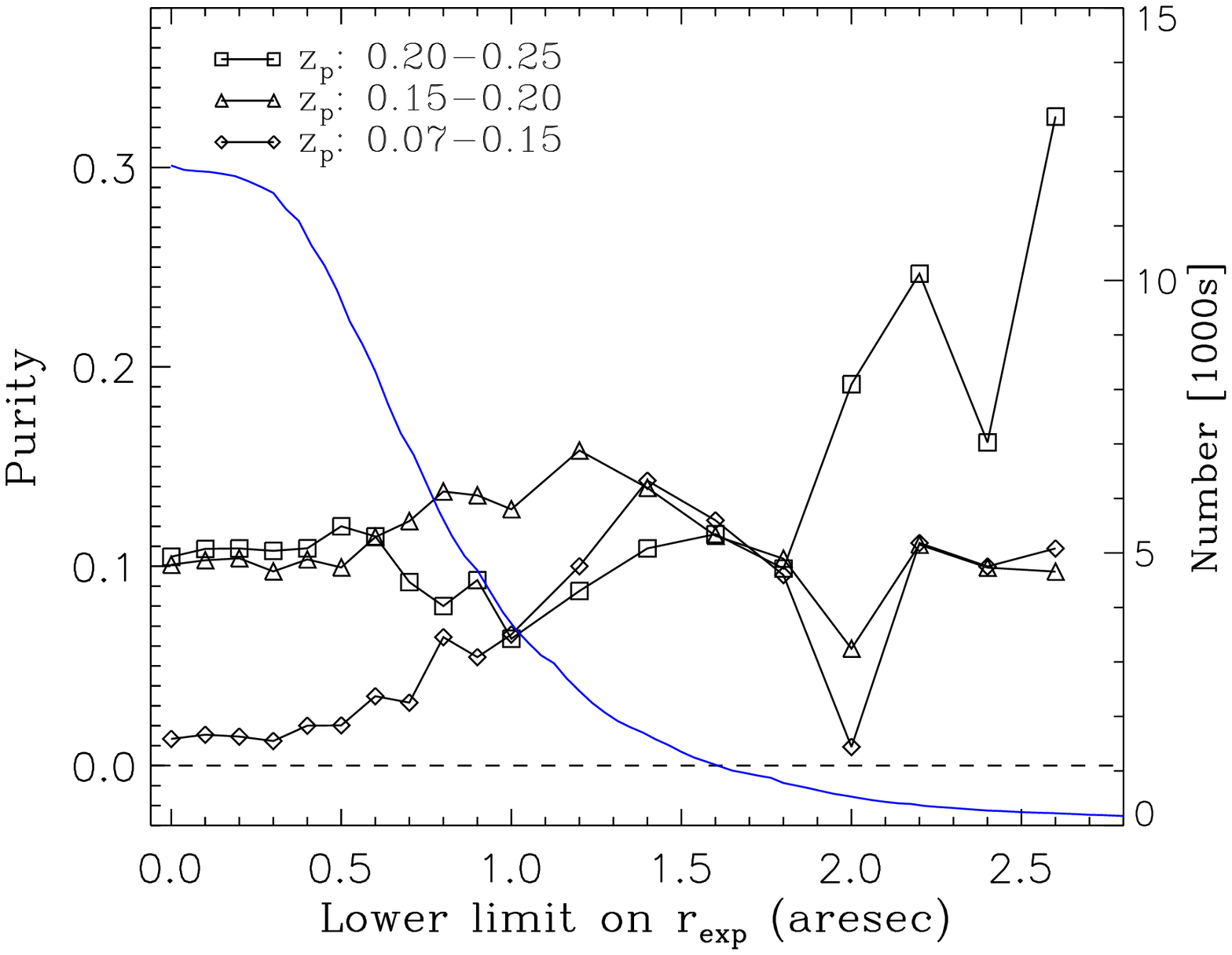}
 	\caption{Top: Purity $P = \Delta N/(\Delta N+ N_{\rm exp})$ of the secondary sample, as a function of a faint magnitude limit. 		Bottom: Purity as a function of a lower size limit on the sample. The smooth (blue) curve in each panel indicates the total 		number of secondaries left in the sample after applying the magnitude cut 
		(with the scale indicated on the right side of the plot).}
 	\label{fig:puritymag}
 \end{figure*}

While {    structural} selection can enhance the SNR of the clustering signal significantly, the purity of the final cut sample, that is the fraction of the sample that is physically associated with the primaries, remains low. In terms of our previously defined quantities, the purity of a cut sample can be defined as the ratio $P = \Delta N/(\Delta N + N_{\rm exp})$. The {    left} panel of Figure~\ref{fig:puritymag} shows the purity of samples produced by a faint magnitude cut. When all but the brightest secondaries are cut out of the sample, the resulting purity is 20\%\ or higher; on the other hand these cuts drastically reduce the size of the sample, and thus the SNR of the clustering measurements. Less severe cuts at $r=$21--22 maximize the SNR, but reduce the purity to 5--15\%\ or less.  In particular, for the lowest redshift primaries, the magnitude cut with the highest clustering SNR produces a final sample with a purity $P\sim$8\%\ . 

The {    right}  panel of Figure~\ref{fig:puritymag} shows the purity of samples produced by a cut on small sizes. Here too, strict cuts on the secondary sample (removing all but the largest galaxies) produce higher purity (25--30\%) but eliminate most of the sample, reducing the overall SNR of the clustering signal. Less strict cuts generally increase the SNR at the expense of purity. The exception is for the lowest redshift primaries, where size cuts around  1.4$\arcsec$  maximize the SNR, while still retaining a purity of almost 15\%\ .

The purity in these {    two} examples, $P\sim$5--15\%\ , is typical for {    all} the single-parameter {    structural} cuts we have considered in this paper. Cuts on two parameters can produce slightly higher purity, as discussed below, but still have $P < $50\%\ . Thus, while {    structural}ly-selected samples are useful for constraining overall satellite abundance, they should be used with caution when, e.g., targeting objects for spectroscopic follow-up.  Even extreme magnitude or size cuts that eliminate most of the sample are relatively ineffective at conclusively identifying individual objects as low-redshift dwarf galaxies, in the absence of spectroscopic information.

\subsection{Two-parameter Cuts}

\begin{figure}
	\includegraphics[width=1.1\columnwidth]{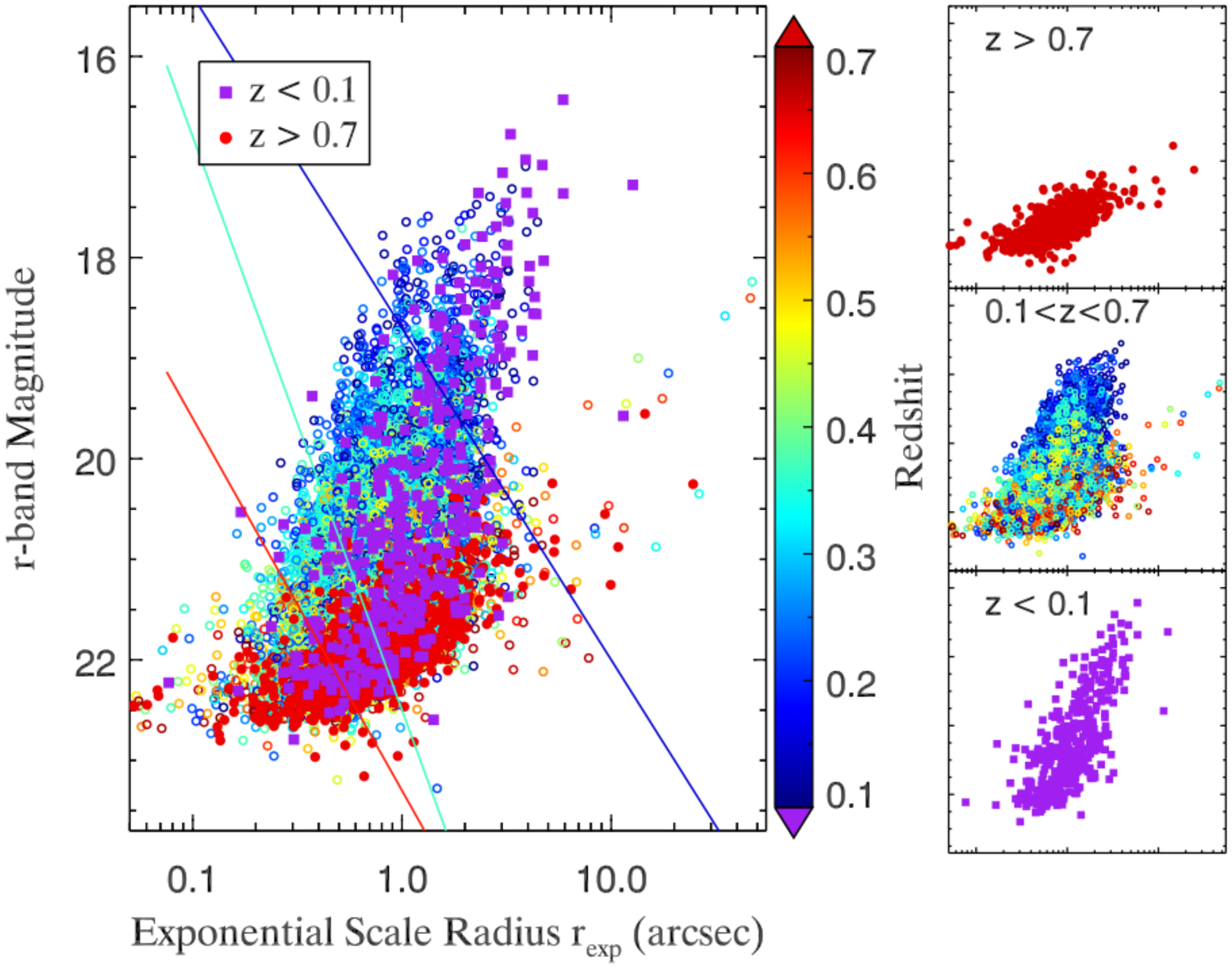}
	\caption{The magnitude-size distribution of the secondary sample.  Points are coloured by redshift (mainly photo-$z$s, corrected with  
	spectroscopic redshifts where they are available). Larger squares and circles indicate the lowest and highest redshift objects, respectively, 
	and are also shown separately in the side panels for clarity. A few data points with very large exponential scale radii are not shown on the plot. 
	Lines indicate the optimal {    structural} cuts in this space, for primary redshift ranges 0.07--0.15 (upper/rightmost line), 0.15--0.20 (middle line) 
	and 0.20--0.25 (bottom/leftmost line). 
	In each case, the {    structural} cut selects galaxies {\it above} the line.}
	\label{fig:sdss_r_vs_size_zcolor}
\end{figure}

\begin{figure}
	\includegraphics[width=1.1\columnwidth]{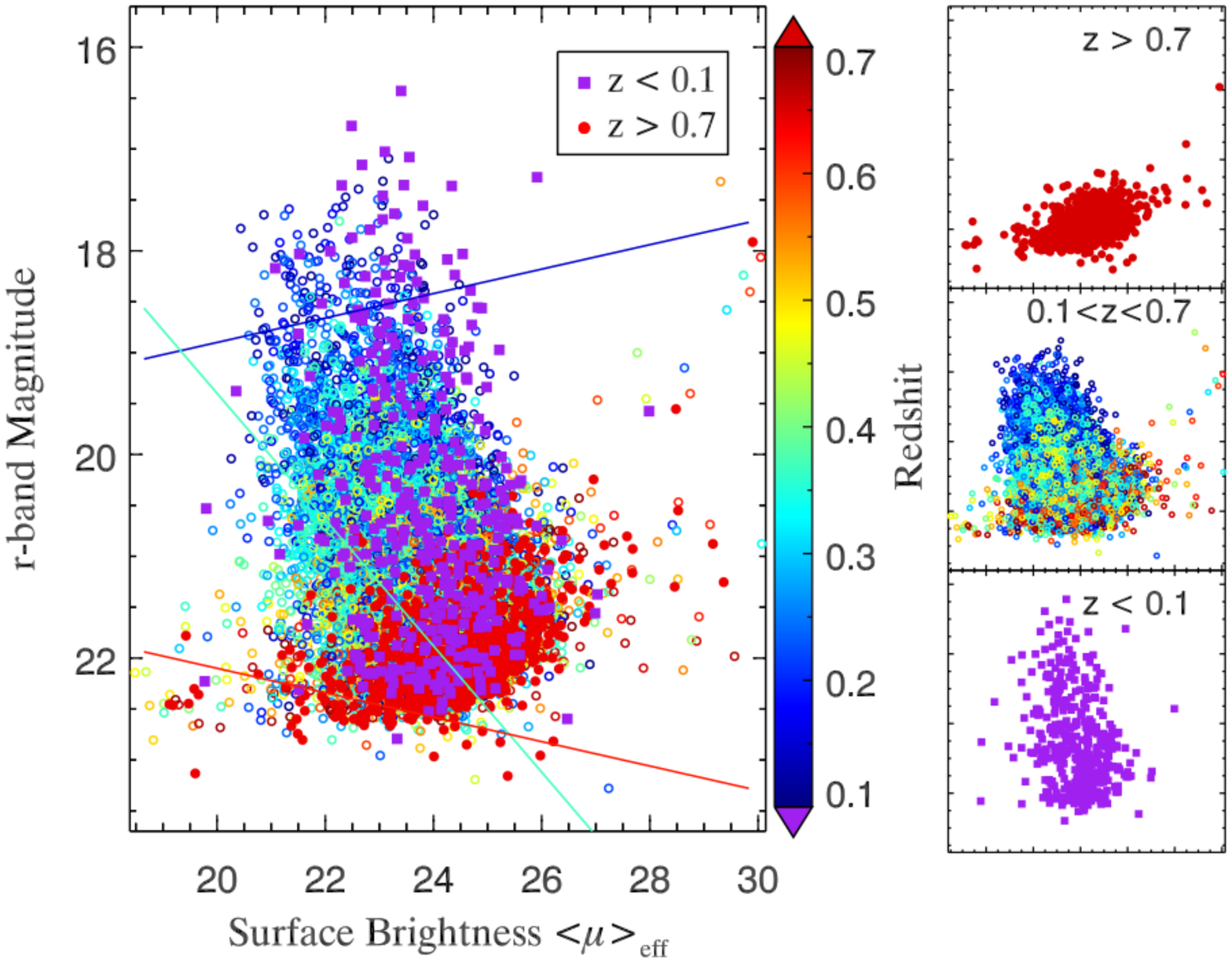}
	\caption{The magnitude-surface brightness distribution of the secondary sample. Points are coloured by redshift, with the larger symbols indicating 
	the highest and lowest redshift objects, as in Figure~\ref{fig:sdss_r_vs_size_zcolor}. 
	Lines indicate the optimal {    structural} cuts in this space, for increasing primary redshift from top to bottom. 
	In each case, the {    structural} cut selects galaxies {\it above} the line.}
	\label{fig:sdss_r_vs_sb_zcolor}
\end{figure}

The distribution of apparent (SDSS $r$-band) magnitude versus size and versus surface brightness for the secondary sample is shown in Figures~\ref{fig:sdss_r_vs_size_zcolor} and~\ref{fig:sdss_r_vs_sb_zcolor}. The colour scale indicates the photometric redshift, while galaxies at $z < 0.1$ or $z > 0.7$ are denoted by larger squares/circles respectively, {    and shown separately in the side panels}. We see that nearby galaxies ($z < 0.1$) are generally larger and lower surface brightness, but that the typical size and surface brightness depend on magnitude. Thus, two-parameter cuts in these planes seem promising for local galaxy selection.

\subsubsection{Size-magnitude Cuts}

First, we consider a size-magnitude cut. After experimenting initially with linear cuts, we found that cuts in log(size) produced slightly higher SNRs. These cuts select objects with $r$-band magnitudes satisfying
\begin{equation}
	r < r_0 + m \log[r_{\rm exp}/1 \arcsec]\,.
\label{eq:szmcut}
\end{equation}
For positive/negative values of $m$, selected objects lie above a line sloping downwards/upwards (since magnitude increases downwards) 
in the magnitude-size space shown in Figure~\ref{fig:sdss_r_vs_size_zcolor}. 
The two free parameters are $r_0$, the intercept of the line at $r_{\rm exp} = 1\arcsec$, and $m$, the slope in $\log(r_{\rm exp})$.

\begin{figure*}
	\includegraphics[width=0.65\columnwidth]{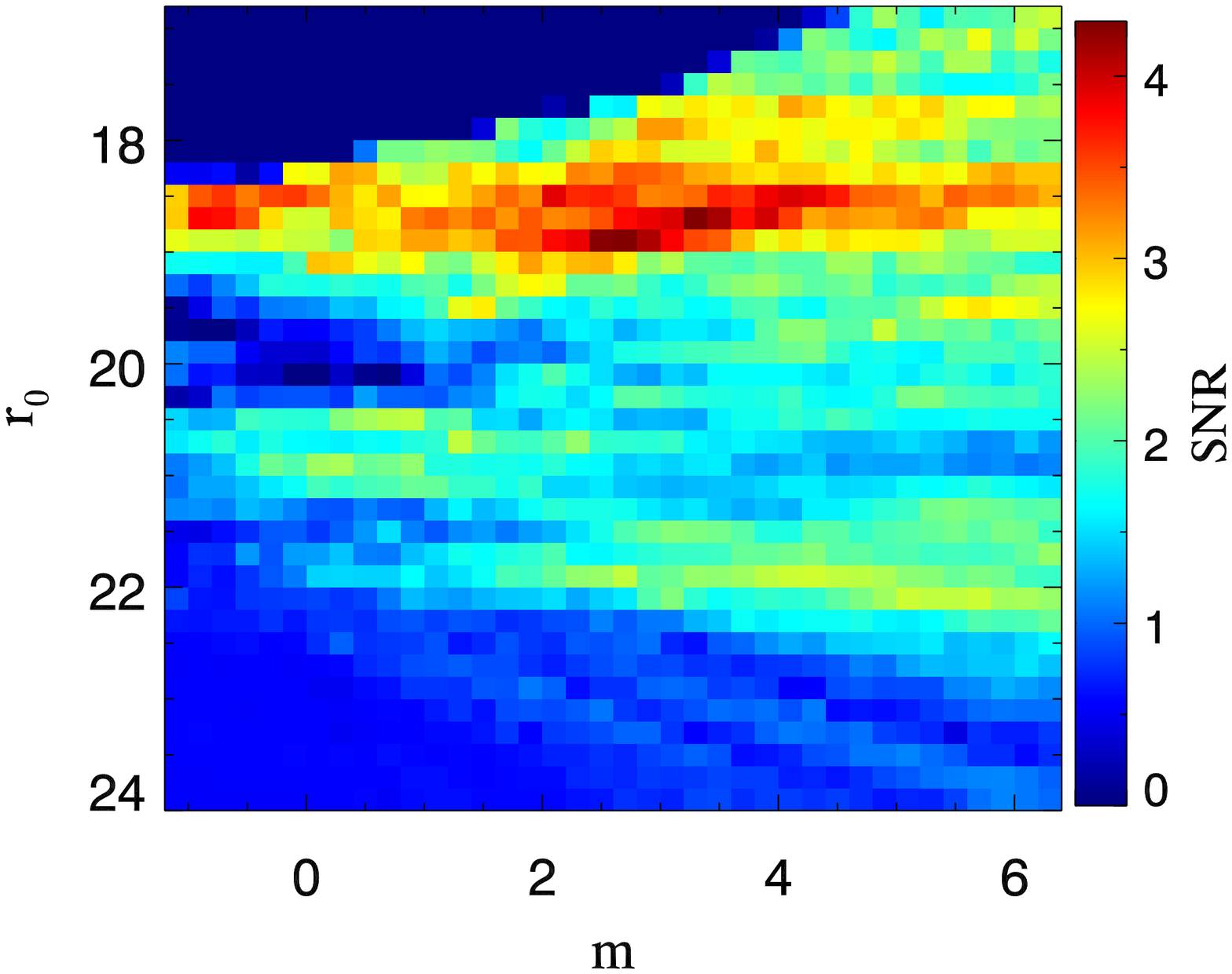}
	\includegraphics[width=0.65\columnwidth]{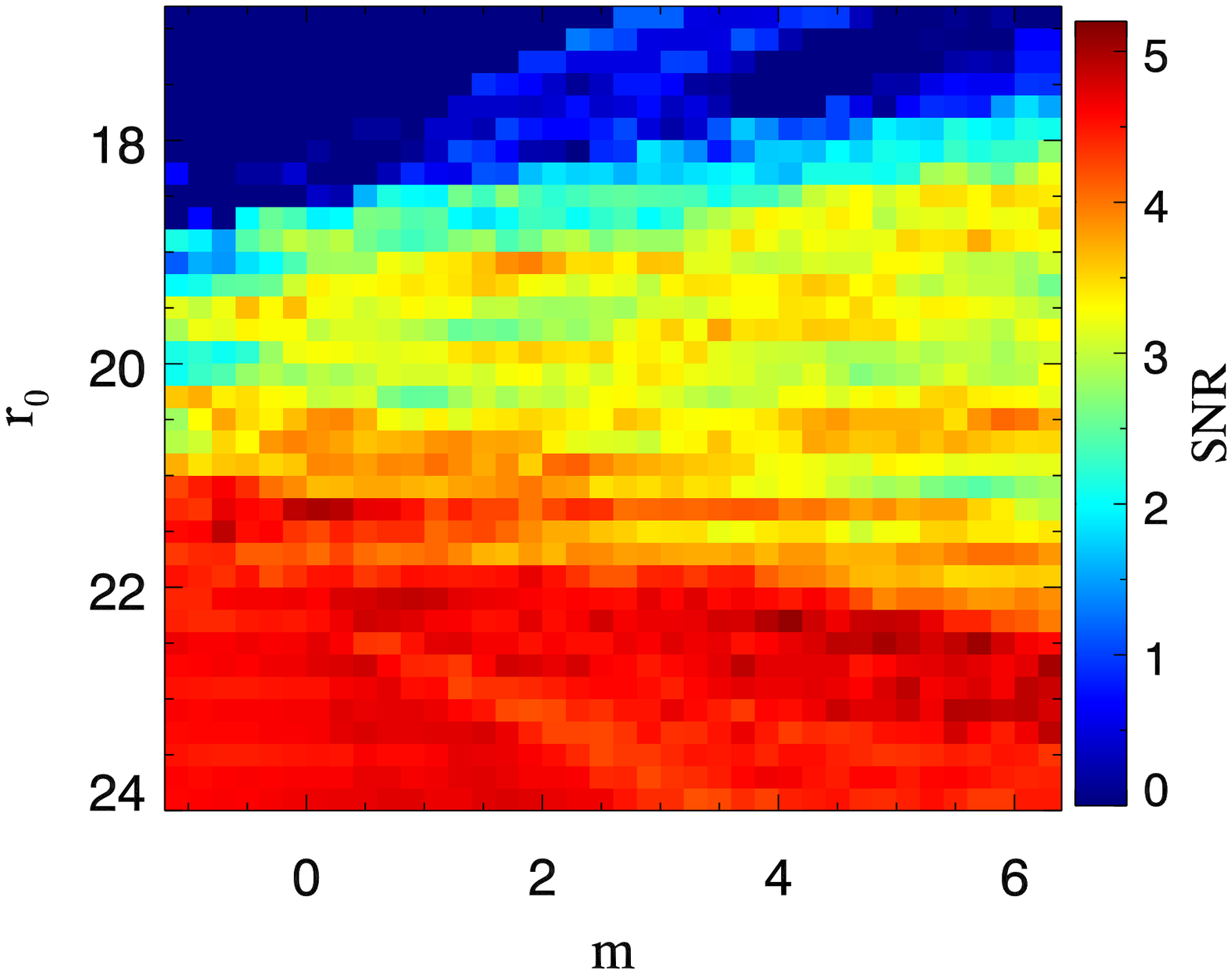}
	\includegraphics[width=0.65\columnwidth]{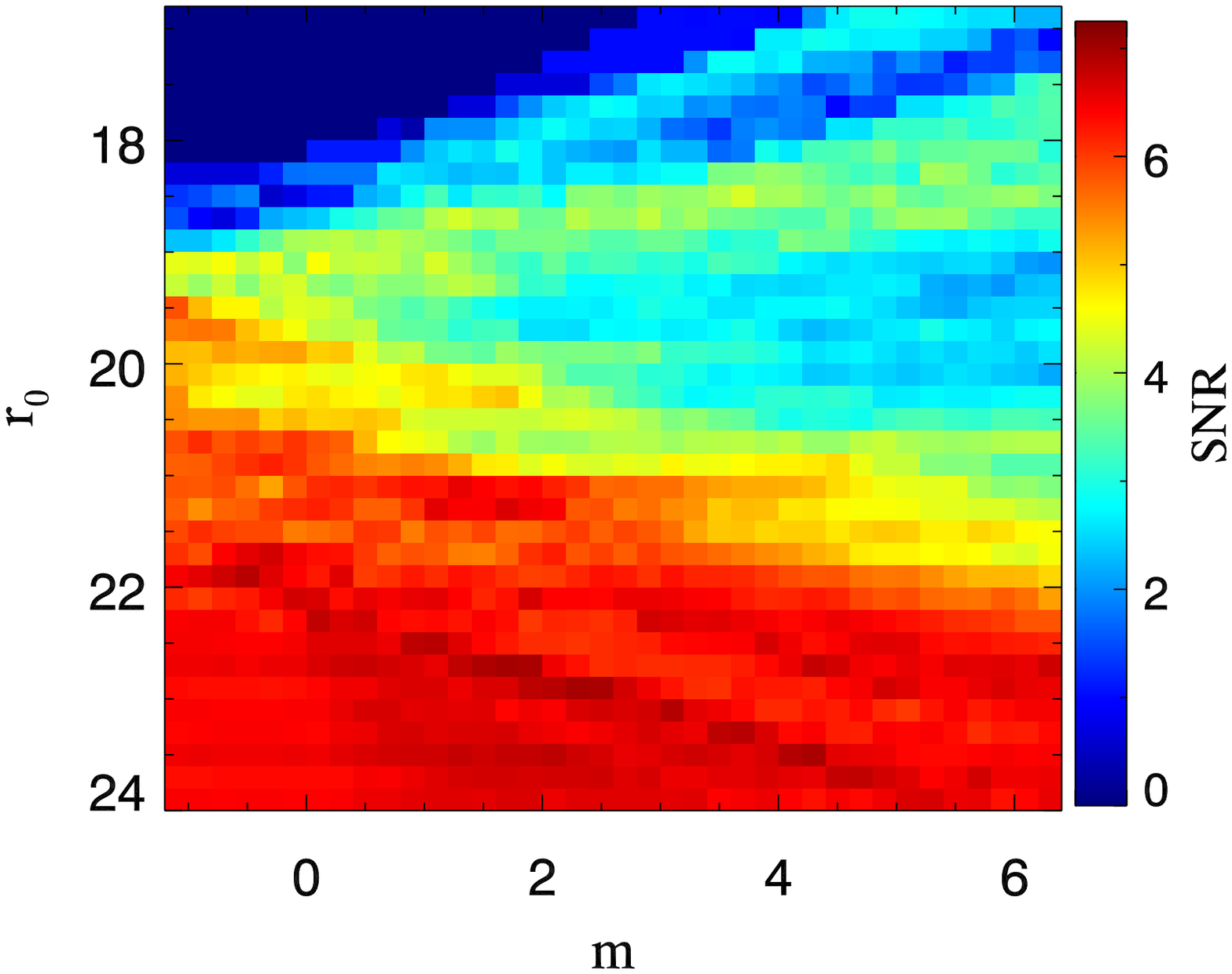}
	\caption{SNR of the clustering measurement as a function of size-magnitude cuts on the secondary sample, parameterized by an intercept $r_0$ 
	and a slope $m$ (Equation~\ref{eq:szmcut}). The three panels are for the three primary redshift ranges 0.07--0.15, 0.15--0.20, and 0.20-0.25, from left to right.}
	\label{fig:snr_slope_m0_rVSsize}
\end{figure*}

In Figure~\ref{fig:snr_slope_m0_rVSsize}, we show the value of our figure of merit (the SNR of the mean excess counts, integrated between projected separations of 50--450 kpc), as a function of the parameters $r_0$ and $m$, for primaries in three redshift ranges. For the lowest redshift primaries, we find that a bright value for $r_0$ and a broad range of positive slopes (from $\sim$2--5) can increase the SNR from $\sim$0 to $\sim$4. This (fairly aggressive) cut removes small and/or faint objects, which are generally farther away. As we move to higher primary redshift, cuts with a fainter value of $r_0$ become optimal, including some with very large slopes $m$. For large values of $m$, these are close to pure size cuts.  Finally at the highest redshift range, faint magnitude cuts produce the highest SNR. In particular, we need to include objects down to $r = 21$--22 or fainter to recover the maximum SNR. The lines on Figure~\ref{fig:sdss_r_vs_size_zcolor} show the location of the best size-magnitude cuts for the three primary redshift ranges. Overall, comparing to our results from Sections~\ref{subsec:nc} and~\ref{subsec:photoz}, we find that size-magnitude cuts only improve the SNR significantly for low-redshift ($z < 0.15$) primaries.

\subsubsection{Surface Brightness-magnitude Cuts}

Next, we consider a cut in surface brightness and magnitude selecting objects with $r$-band magnitudes
\begin{equation}
r<r_0 +m (\mu -25)
\label{eq:sbmcut}
\end{equation}
where $r_0$, the intercept at $\mu=25$, and $m$, the slope, are the two free parameters, and surface brightnesses are all $\langle\mu\rangle_{\rm eff}$. 

\begin{figure*}
	\includegraphics[width=0.65\columnwidth]{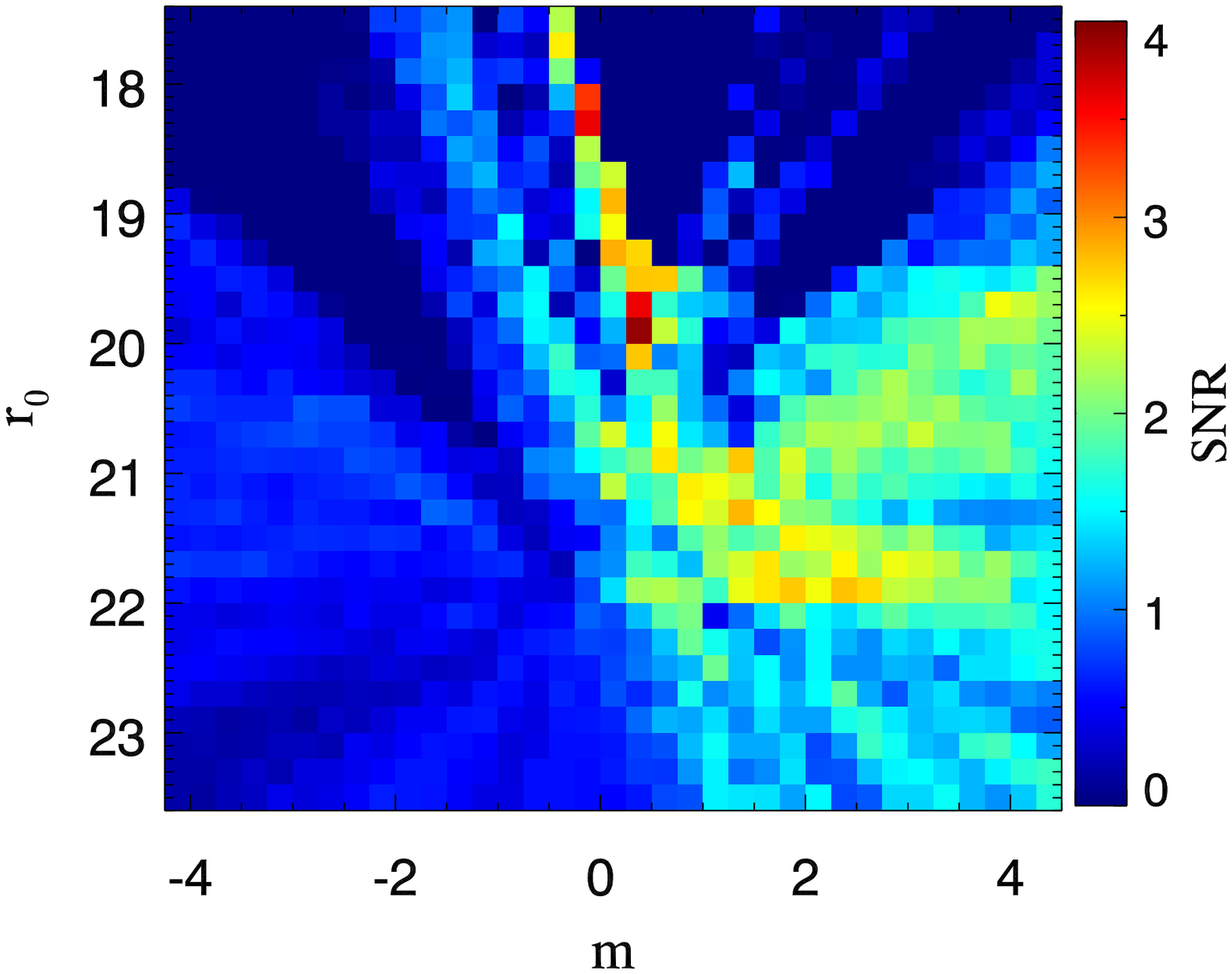}
	\includegraphics[width=0.65\columnwidth]{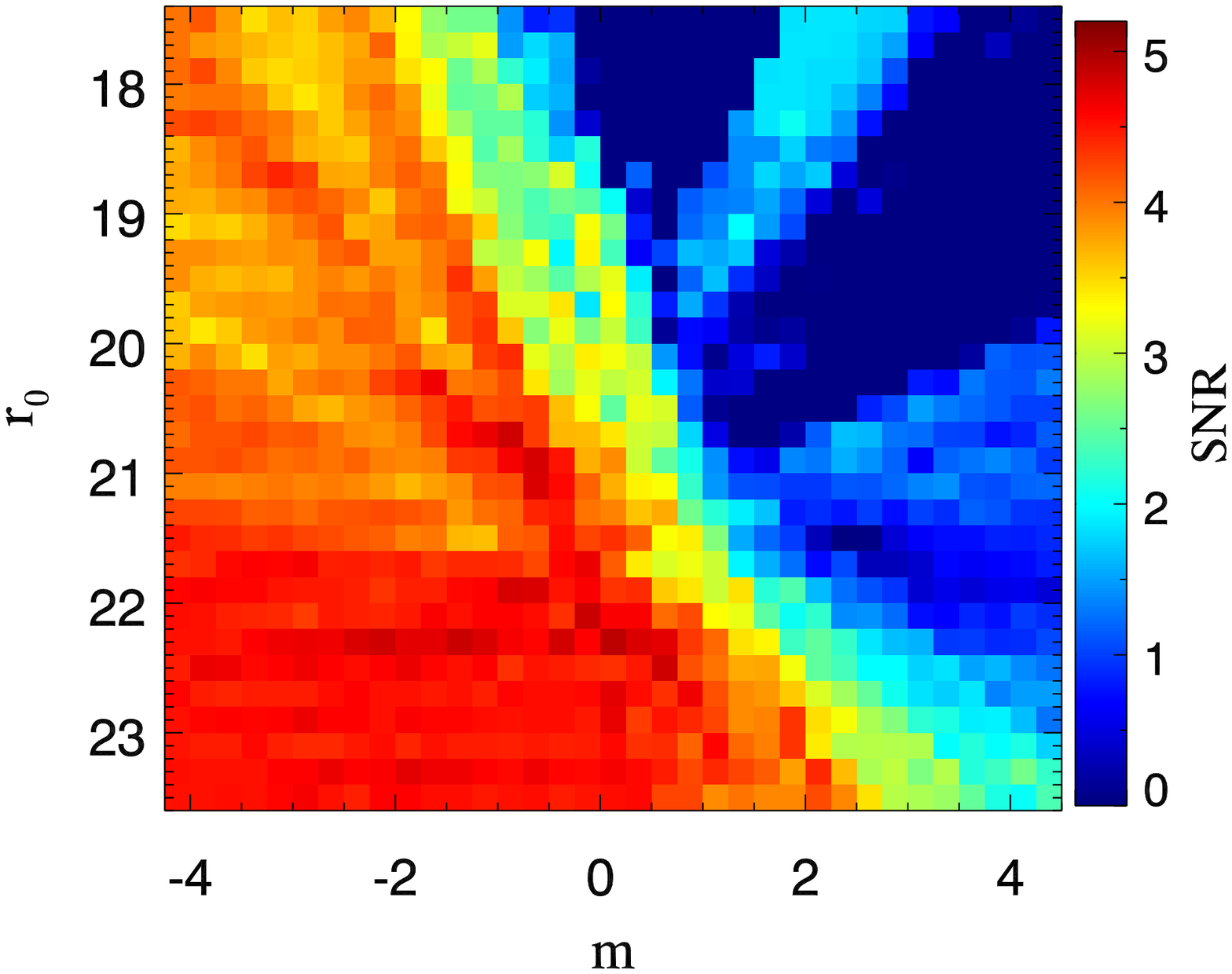}
	\includegraphics[width=0.65\columnwidth]{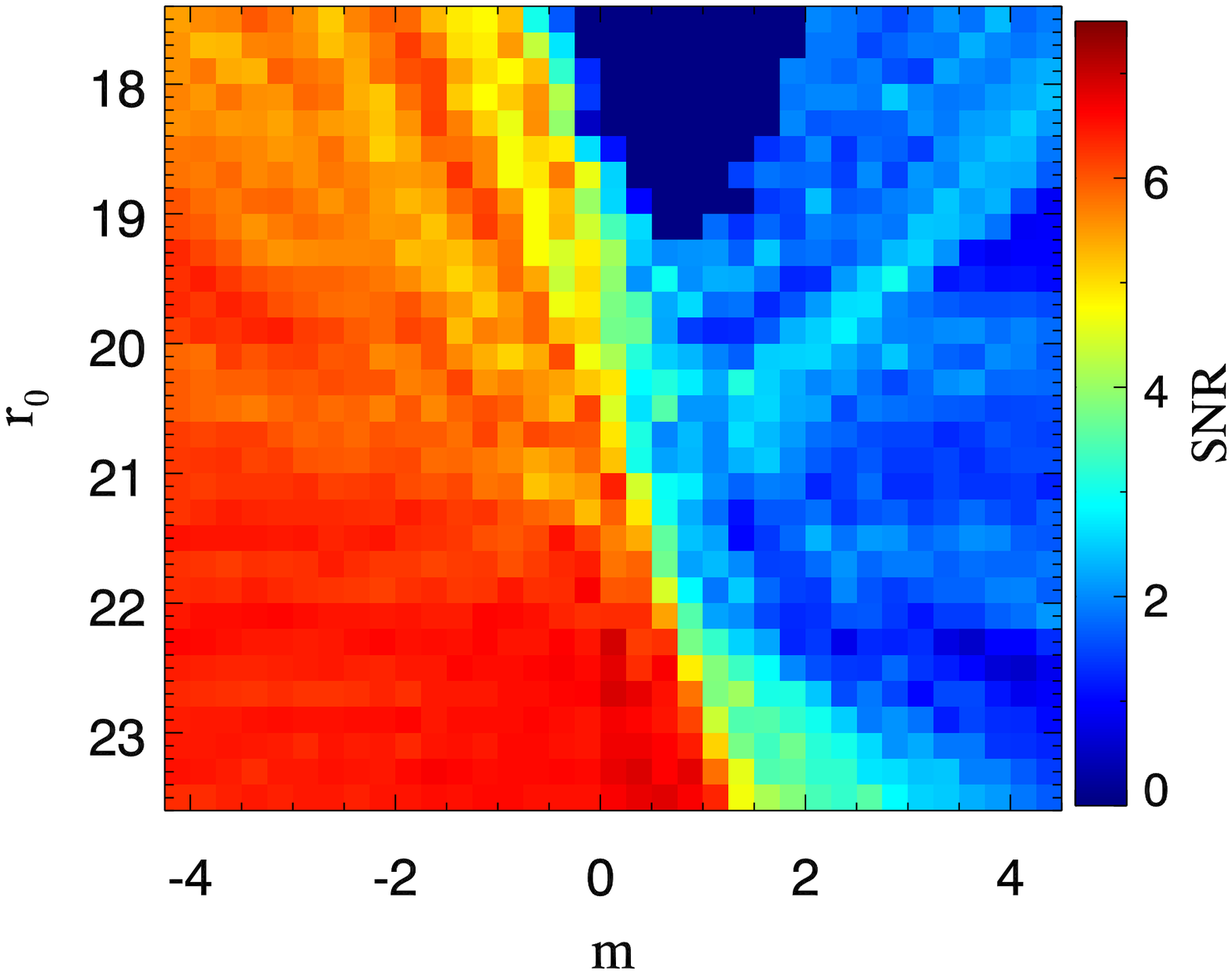}
	\caption{SNR of the clustering measurement as a function of surface brightness-magnitude cuts on the secondary sample, parameterized by 
	an intercept $r_0$ and a slope $m$ (Equation~\ref{eq:sbmcut}). Panels are as in figure \ref{fig:snr_slope_m0_rVSsize}.}
	\label{fig:snr_slope_m0_rVSsb}
\end{figure*}

The results of this cut are shown in Figure~\ref{fig:snr_slope_m0_rVSsb}. As in the previous figure, we see an initial pattern for low-redshift primaries (top left panel), that gradually changes as we move to higher primary redshift. At the lowest redshifts, this cut is relatively ineffective, except for one or two specific points in parameter space, which may simply reflect noise in the sampling or the clustering measurement. As the primary redshift limit increases, we find that cuts at fairly faint $r_0$ with slopes close to $m=0$ (i.e.~pure magnitude cuts) do best. Finally, for the highest redshift limit, any cut with a negative slope seems to work well. The lines on Figure~\ref{fig:sdss_r_vs_sb_zcolor} show the location of the best cuts for the three primary redshift ranges.

\subsection{Optimal {    Structural} Cuts}
\label{subsec:optimal}

Table~\ref{tbl:optimal} lists the optimal parameter choices (i.e.~those that maximize our figure of merit, the SNR of the clustering measurement) for the (log) size-magnitude cuts  (first six columns) and the surface brightness-magnitude cuts (last six columns). For comparison, in the last two rows of each section of the table, we also list the corresponding SNRs for the secondary catalogue with no cuts (SNR$_{\rm nc}$), or with photo-$z$ cuts around each primary (SNR$_{\rm pz}$). These SNRs were shown previously in the lower panels of Figures \ref{fig:ns_z015_z020_z025} and \ref{fig:ns_zcut_z015_z020_z025}. 

\begin{table*}
	\centering		
	\caption{Optimal values for cuts in size and magnitude (left six columns) and surface brightness and magnitude (right six columns). SNRs for clustering measurements with no cuts {    (SNR$_{\rm nc}$)} and with photo-$z$ cuts {   (SNR$_{\rm pz}$)} are given for comparison.} 
	\begin{tabular}{cccccc|c|cccccc}
		\hline
		Redshift Range & $m$ & $r_0$ & SNR & SNR$_{\rm nc}$ & SNR$_{\rm pz}$ & & Redshift Range & $m$ & $r_0$ & SNR & SNR$_{\rm nc}$ & SNR$_{\rm pz}$\\
		\hline\\
                	0.07--0.15 & 3.3 & 18.7 & 4.3 & 0.6 & 5.5 & &  0.07--0.15 & -0.12 & 18.3 & 3.9 & 0.6 & 5.5 \\
                	\\
                	0.15--0.20 & 5.7 & 22.5 & 5.4 & 4.5 & 6.9 & &  0.15--0.20 & 0.62 & 22.5 & 5.1 & 4.5 & 6.9 \\ 
                	\\
                	0.20--0.25 & 3.7 & 23.3 & 6.9 & 6.4 & 9.8 & &  0.20--0.25 & 0.12 & 22.7 & 6.7 & 6.4 & 9.8 \\
                	\\
		\hline
	\end{tabular}
	\label{tbl:optimal}
\end{table*}

\begin{figure}
	\includegraphics[width=\columnwidth]{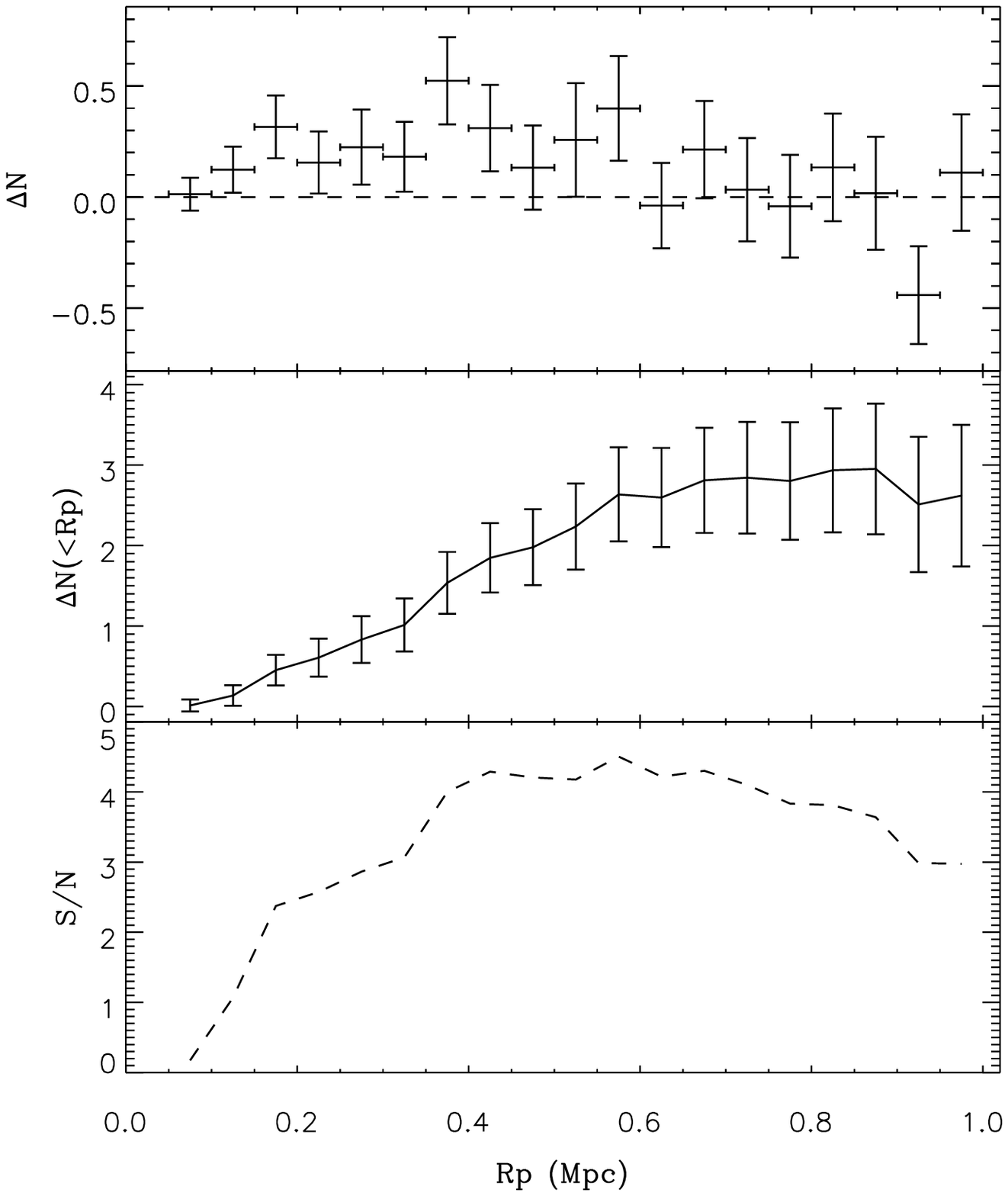}
	\caption{The clustering signal between primaries in the redshift range $z=$0.07--0.15 and the secondary sample after an optimal size-magnitude cut has been applied. }
	\label{fig:cluster_z150_bestMorph}
\end{figure}

Figure~\ref{fig:cluster_z150_bestMorph} shows the clustering signal around the lowest-redshift primaries ($z < 0.15$), for the best of the {    structural} cuts we have tested, a cut in (log) size and magnitude with the parameters listed in Table~\ref{tbl:optimal}. The SNR reaches a value of 4.3 at 450 kpc, compared to 5.5 for the photo-$z$-selected sample (Figure~\ref{fig:ns_zcut_z015_z020_z025}), or 0.6 for the uncut secondary sample (Figure~\ref{fig:ns_z015_z020_z025}). Thus, we recover about 80\%\ of the maximum SNR obtainable with COSMOS-quality photo-$z$s. We can also calculate the purity of the cut sample, $P = \Delta N/(\Delta N + N_{\rm exp})$. For the optimal size-magnitude cut this is relatively high, $P = 0.34$, so more than a third of selected objects are genuine satellites.

\begin{figure}
	\includegraphics[width=\columnwidth]{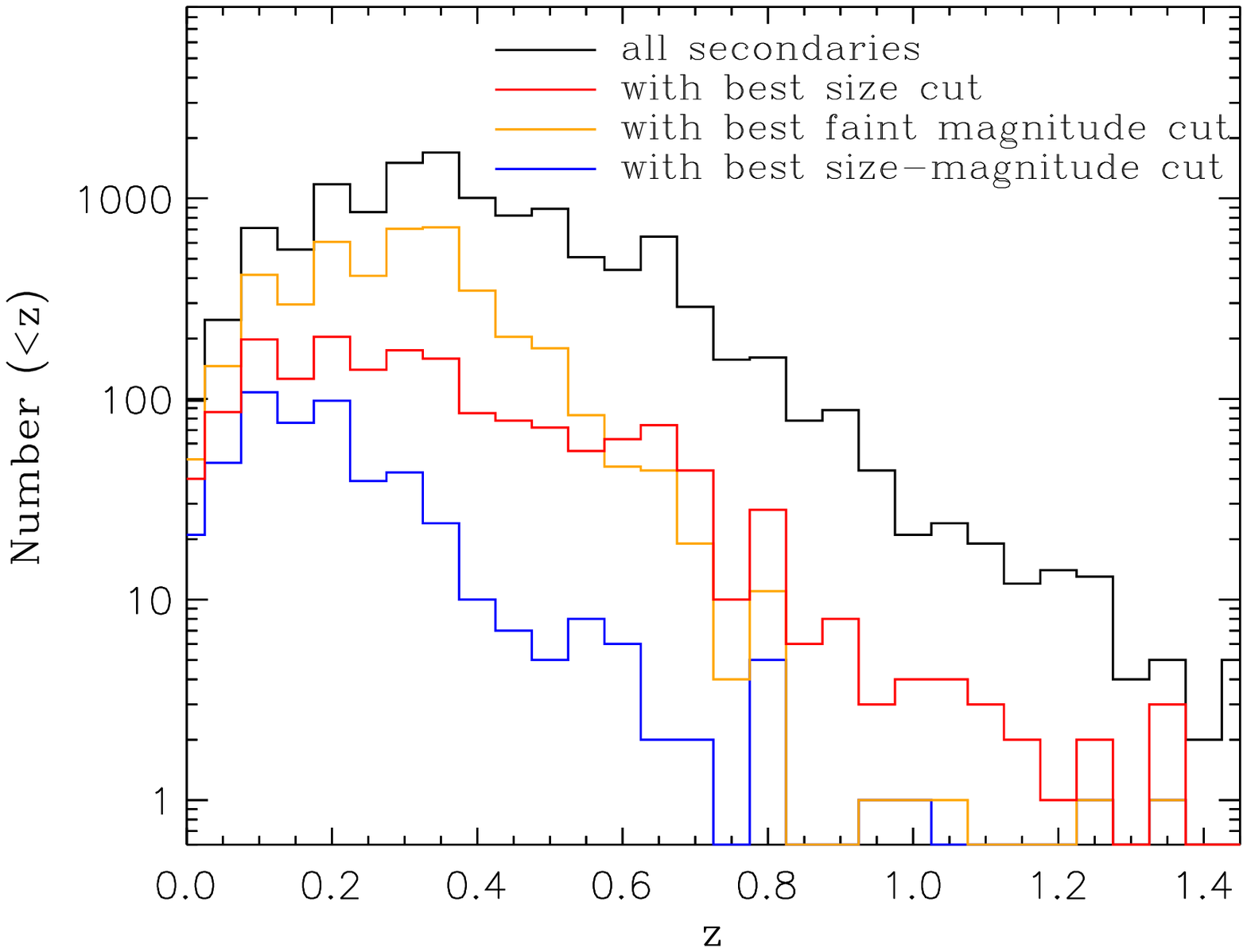}
	\caption{The redshift (mainly photo-$z$) distributions of the entire secondary sample, and secondary samples after the best cuts in magnitude, in size, 
	or in both size and magnitude have been applied.}
	\label{fig:zdist}
\end{figure}

The net effect of the {    structural} cuts on the redshift distribution of the secondaries can be seen by comparing the photo-$z$s of the uncut and cut samples. Figure~\ref{fig:zdist} shows these distributions for the entire secondary sample, and after applying best single-parameter cuts in magnitude or size, or the best size-magnitude cut (our `optimal' cut). We see that a size cut on its own is of limited use, as it reduces the size of the sample but not the shape of the redshift distribution, except perhaps at very low redshift. A cut in magnitude is more effective, reducing the number of objects at $z > 0.4$, and eliminating most objects beyond z$\,\sim\,$0.6--0.8. The optimal size-magnitude cut is most effective, however, eliminating most objects beyond $z = 0.4$, and shifting the peak of the redshift distribution from $z=0.35$ to $z=0.1$.

The completeness of our cut sample, relative to a photo-$z$ selected one, is a little less clear. On the one hand, applying the optimal (size-magnitude) cut to the secondary catalogue reduces the number of objects with photo-$z$s below 0.15 to 18\%\ of the uncut number, suggesting our completeness should be $\sim\,$20\%\ or less. The best magnitude or size cuts reduce the sample size by similar factors. On the other hand, comparing Figures~\ref{fig:cluster_z150_bestMorph} and \ref{fig:ns_zcut_z015_z020_z025}, we see the excess counts reach a value of $\Delta N \sim 2$ at 450 kpc, or $\Delta N \sim 3$ at large radii in the {    structurally selected secondary sample with the optimal cut}, versus $\Delta N = 3$ or $\Delta N = 4$, respectively, in the photo-$z$ selected sample. This suggests that the cut sample contains 66--75\%\ of the {    true} satellites in the photo-$z$ selected one (with an uncertainty of about 20\%\ on that fraction). One possible resolution to this puzzle is if the photo-$z$ selected sample is itself incomplete, due to missing photo-$z$s, catastrophic failures, or other problems. In this case, the amplitude of the clustering signal in Figure~\ref{fig:ns_zcut_z015_z020_z025} would be an underestimate of the true signal. At the moment, we will content ourselves with comparing the {\it relative} 
performance of {    structural} selection and photo-$z$ selection, and leave a discussion of absolute performance and completeness to future work.

Overall, we conclude that at low redshift ($z < 0.15$), {    structural} selection can be very effective, recovering $\sim$80\%\ of the clustering signal obtainable with high quality photo-$z$s, with reasonable completeness and purity  ($\sim$66\% and 33\%\ respectively, albeit with some uncertainty in the absolute completeness). This result is particularly impressive, given that we have considered only simple cuts, defined either by a single limit in magnitude, size or surface brightness, or by a linear relation between magnitude and log(size) or between magnitude and surface brightness. Furthermore, our cuts are based on relatively shallow SDSS photometry, whereas the COSMOS photo-$z$s are based on 30 bands of photometry, most of it from much deeper and/or higher-resolution imaging.

At higher redshift, it is worth noting that these simple {    structural} cuts are {\it not} as effective. The highest SNRs we achieve, 5.4 for $z < 0.15$--0.20 and 6.9 for $z =$ 0.20--0.25, are only slightly higher than those obtained without any cuts on the secondary catalog (cf.~Figure~\ref{fig:ns_z015_z020_z025}), indicating that we have not succeeded in separating foreground and background galaxies very effectively at these distances. 

\subsection{Completeness and Bias in Other Properties}

\begin{figure}
 	\includegraphics[width=\columnwidth]{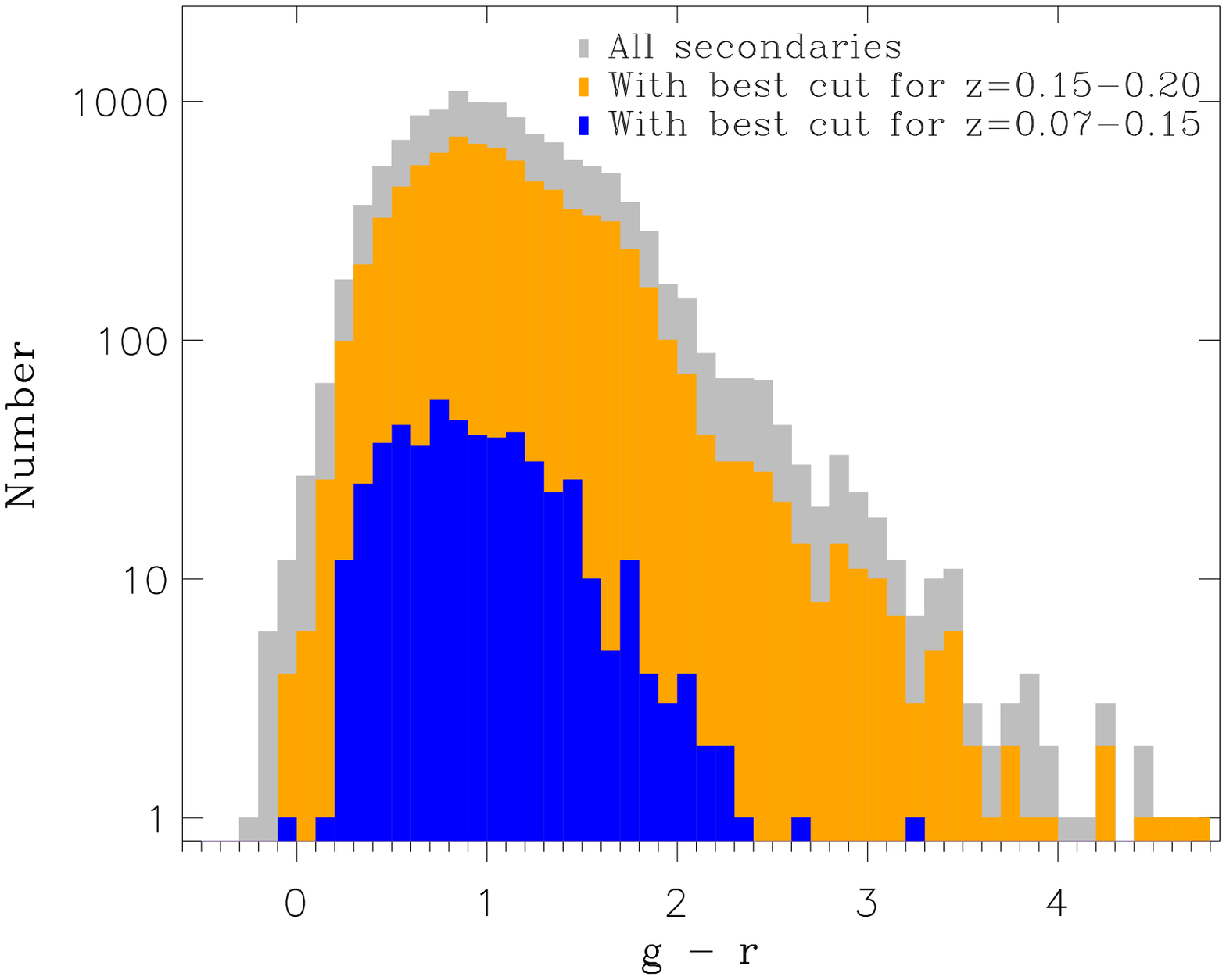}
 	\caption{From top to bottom, the (SDSS) $g-r$ colour distribution of the entire secondary sample (grey), the colour distribution of the sample after the optimal size-		magnitude cut for the redshift range 0.15--0.2 is applied (orange), and the distribution for the sample after the optimal size-magnitude cut 
	for the redshift range 0.07--0.15 is applied (blue).}
 	\label{fig:clustering_vs_color_withbest}
 \end{figure}

While we have shown {    structural} selection can be effective in preferentially selecting faint satellites around nearby galaxies, even out to redshift $z\sim0.15$, one potential concern is the completeness of such samples, and any biases that {    structural} selection may introduce in other satellite properties. In particular, since red and blue galaxies differ in structure, we might expect {    structural} selection to bias the colour distribution of the final samples. To test this possibility,  Figure~\ref{fig:clustering_vs_color_withbest} compares the (SDSS) $g-r$ colour distribution for the whole secondary sample, and the distributions after two of the optimal size-magnitude cuts are applied. The distributions look remarkably similar, modulo an overall scaling, although the cuts do shift the mean colour slightly to the blue (from $\left<g-r\right> = 1.11$ for the whole sample to  $\left<g-r\right> = 1.10$ after the optimal size-magnitude cut for the redshift range 0.15--0.2 is applied, or $\left<g-r\right> = 1.03$ after the optimal size-magnitude cut for the redshift range 0.07--0.15 is applied).

\begin{figure}
	\includegraphics[width=0.9\columnwidth]{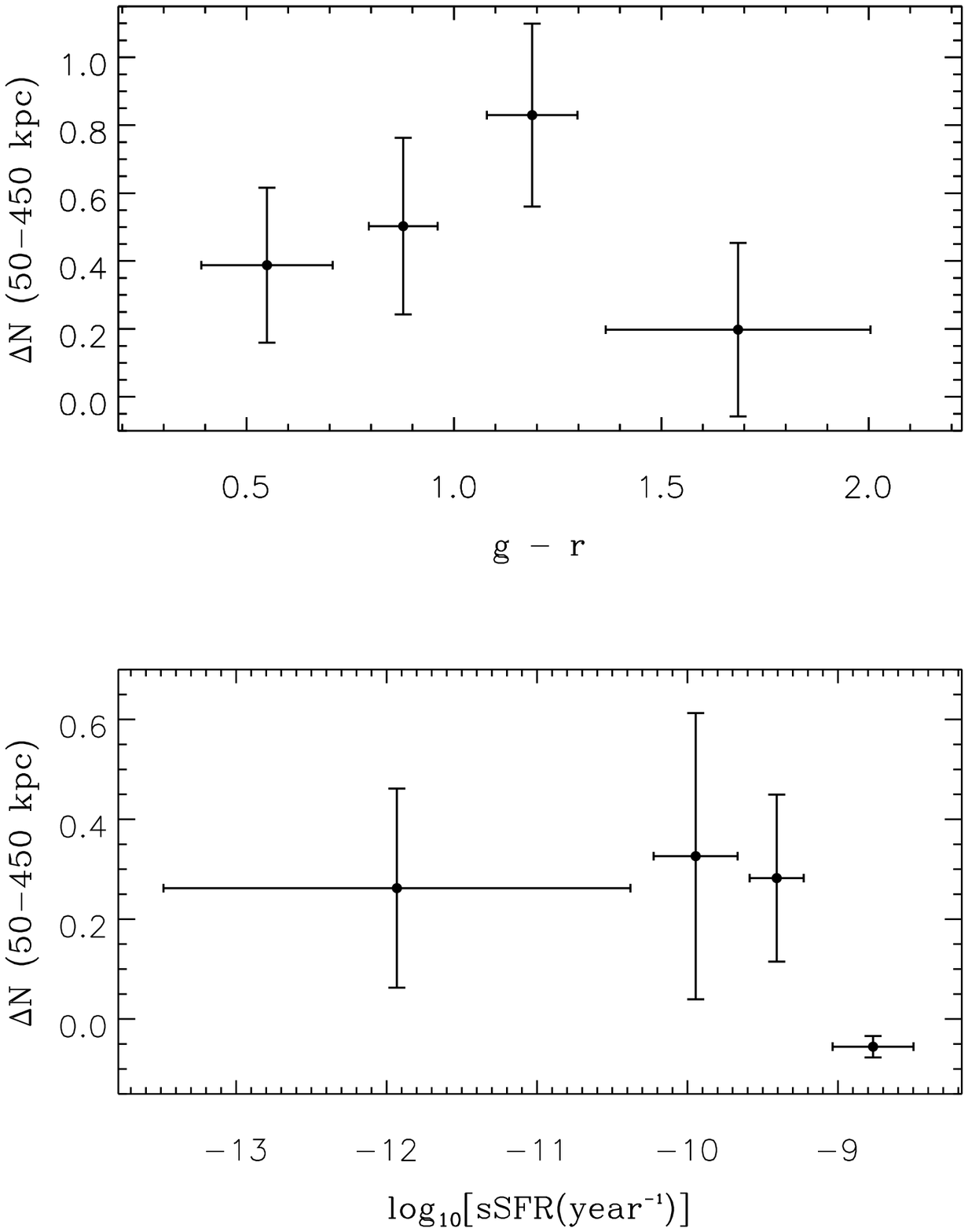} 
	\caption{{\it Top panel:} excess counts $\Delta N$ as a function of secondary colour, around primaries in the redshift range $z=0.07$--0.15. 
	{\it Bottom Panel:} excess counts as a function of specific star formation rate, for the same redshift range. In both panels, 
	the optimal size-magnitude cut for this redshift range has been applied to the secondary sample.}
	\label{fig:clustering_vs_csgr}
\end{figure}

In fact, a significant part of the clustering signal comes from galaxies with blue or intermediate colours. 
The top panel of Figure~\ref{fig:clustering_vs_csgr} shows the excess counts integrated 
from 50 to 450 kpc around primaries in the redshift range $z=0.07$--0.15, as a function of satellite colour. As before, the optimal size-magnitude cut for this redshift range
has been applied to the secondary sample. We see roughly equal signals from all three blue bins, but less signal 
for the reddest bin (albeit with large uncertainties). On the other hand, our satellites are not necessarily forming stars rapidly. The bottom panel of Figure~\ref{fig:clustering_vs_csgr} shows 
the clustering signal for secondaries binned by specific star formation rate (SSFR, as derived in the COSMOS 2015 catalog -- cf.~\citealt{Laigle16}). Here we 
see that passive galaxies are generally more clustered than active ones. This suggests that the pattern of clustering with colour seen in the top panel may be 
a result of the redshift distribution of the secondary sample, rather than a dependence on rest-frame colour. In some applications, colour cuts might provide a 
useful addition to {    structural} cuts in selecting satellites, albeit with significant implications for completeness.

\section{Other Morphological Distance Indicators}
\label{sec:7}

Finally, while working with the COSMOS catalog, we have noted (and have had pointed out to us) many individual galaxies that appear to be nearby from their detailed morphology, showing features such as multiple point sources in the Hubble Space Telescope (HST) imaging. Although it is slightly tangential to our main argument, in this section we will briefly consider the use of these detailed morphological features to estimate distances to very nearby dwarfs.  

\subsection{Serendipitous Discoveries and their Redshifts}

Over the years, close examination of COSMOS HST images has revealed a number of galaxies that appear to be resolved, partially resolved, or otherwise unusual. Through visual examination, we have divided these serendipitous discoveries into seven rough classes:
\begin{enumerate}
\item {\bf Class 1} objects contain many clearly recognizable point sources, which together account for a significant fraction of their light. The implication is that they are close enough to be resolved into regions dominated by individual bright stars in the COSMOS ACS images (which have a resolution of approximately {    $0.095\arcsec$ in F814W -- cf.~\citealt{Koekemoer2007})}. 
\item {\bf Class 2} may be resolved or partially resolved into point sources, but are less distinct than Class 1.
\item {\bf Class 3} objects appear to be high surface-brightness galaxies at larger distances, but still close enough to have visible \mbox{H\,{\sc ii}} regions and the like.
\item {\bf Class 4} objects are large and extremely LSB, suggesting some or all may be local LSB dwarfs.
\item {\bf Class 5} objects appear to be distant galaxies whose light is significantly contaminated by a single bright galactic star superposed on the galaxy.
\item {\bf Class 6} objects are LSB galaxies with a few superposed point sources that may or may not be foreground galactic stars.
\item {\bf Class 7} includes all other strange or unusual objects that look like they might be nearby.
\end{enumerate}
Class 1 appears to be complete, in the sense that an examination of bright, low-redshift objects in the photo-$z$ catalog reveals no other similar objects that were not already discovered serendipitously. Class 2 appears to be fairly complete as well, although it may be missing some similar objects. The other classes are very incomplete, though enough objects are known in each to provide a representative sample.

\begin{figure}
	\vspace{-14mm}
        \includegraphics[width=0.99\columnwidth]{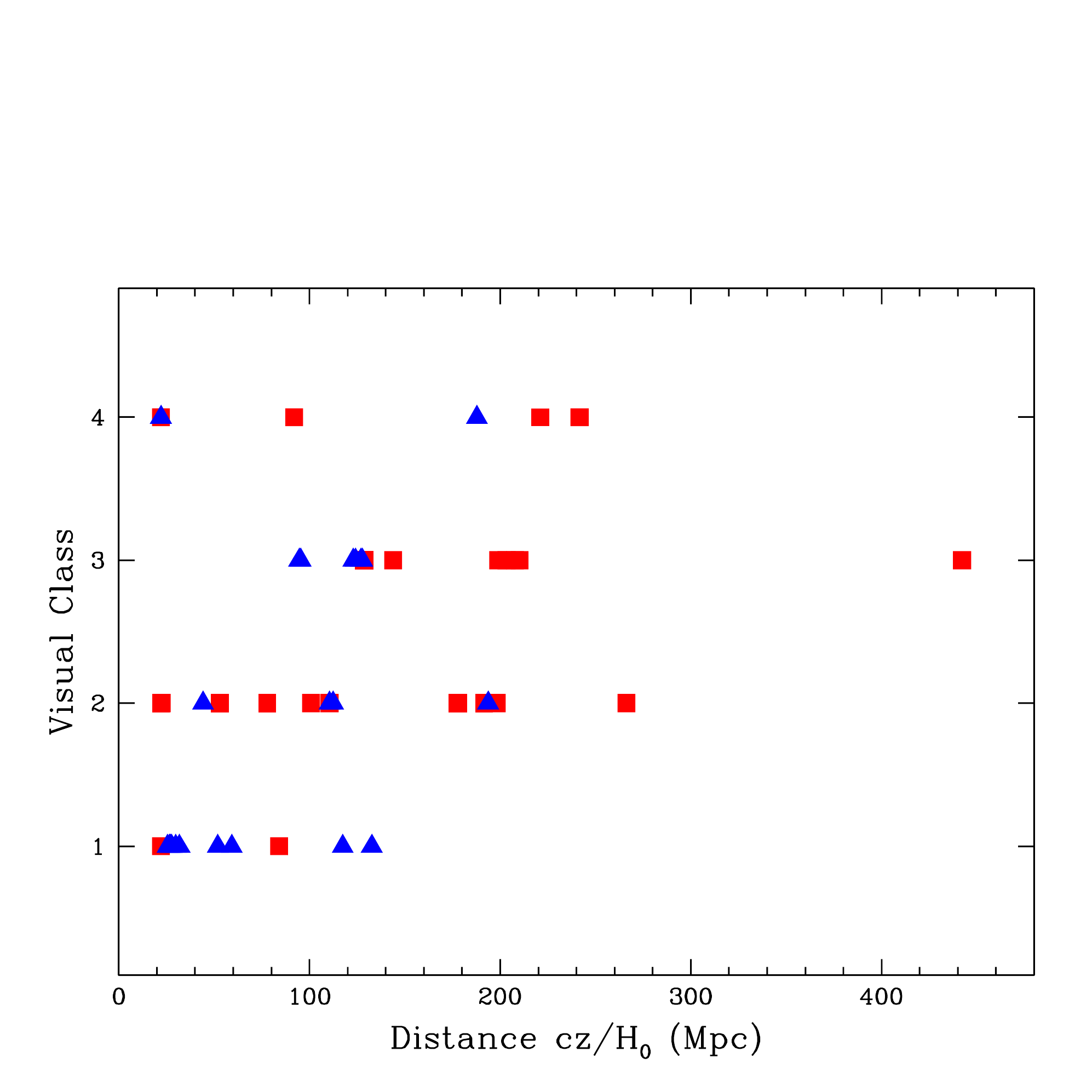}
    \caption{Visual class versus distance, inferred from photometric (red squares) or spectroscopic (blue triangles) redshift, for the serendipitous sample.}
    \label{fig:classvsdist}
\end{figure}

\begin{figure}
	\vspace{2mm}
	 \includegraphics[width=\columnwidth]{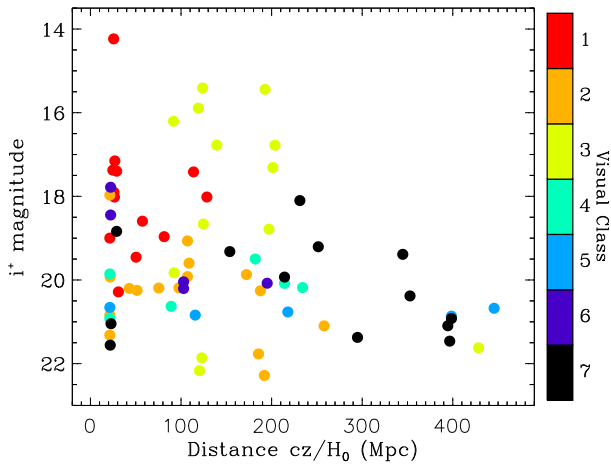}
    	\caption{Apparent magnitude versus distance for the serendipitous sample.}
        	\label{fig:magvsdist}
\end{figure}

Given the COSMOS photo-$z$s are accurate at the percent level, even at low redshift, as discussed in section~\ref{sec:2}, 
we can use them to verify the robustness of this visual classification. Figures~\ref{fig:classvsdist} and \ref{fig:magvsdist} show the visual class and magnitude respectively, plotted versus distance inferred from the (photometric or spectroscopic) redshift.\footnote{We note that in a few cases, objects in the serendipitous sample had neither a spectroscopic redshift, nor a single converged photo-$z$ from template fitting. In these cases we took the midpoint between the 68\%\ confidence limits as the estimated photo-$z$.} For Classes 1--3, we see that visual classification is surprisingly effective. All objects classified visually as being clearly nearby (Class 1) lie at distances less than $D=130$ Mpc, and all but two are at $D < 80$ Mpc. The less certain Class 2 objects are also fairly local, but lie at distances up to 260 Mpc. Class 3 objects, which appear to be more distant visually, generally are further away, with minimum distances of 90 Mpc. The other classes consist of objects whose distances are harder to estimate, or may be incorrect due to foreground contamination; as expected, their photo-$z$s indicate that they lie at a wide range of distances (Classes 5--7 are possibly contaminated and/or confusing objects, so we do not include them in Figure~\ref{fig:classvsdist}). 

The distribution of serendipitous identifications with distance and magnitude (Figure~\ref{fig:magvsdist}) also sheds some light on the net outcome of visual classification. Bright objects within 100 Mpc are generally assigned Class 1; bright objects at larger distances are generally assigned Class 3; Class 2 objects are generally fainter and lie at a range of distances, while the other classes, similarly, are faint and spread over a range of distances. We note that in some cases, the success of visual classification is circular; the objects in the serendipitous catalog come from many different sources, and some were flagged as having low photo-$z$s before they were examined visually. The majority of the serendipitous discoveries were identified directly in the HST imaging {\it before} their photo-$z$ was checked, however, so overall we can confirm that visual classification works fairly well, even in the absence of other information.

From these figures, we conclude that visual classification of images with HST resolution can reliably identify bright ($i^{+} < 19$) local galaxies out to distances of $\sim$100 Mpc, and can identify some fainter ($i^{+} = 19$--21) galaxies out to $\sim$250 Mpc. The COSMOS field alone has more than a dozen of each, in an area of less than 2 square degrees. This has interesting implications for future wide-field, space-based imaging surveys.  Surveys such as those planned with Euclid\footnote{http://sci.esa.int/euclid} and WFIRST\footnote{https://wfirst.gsfc.nasa.gov} can expect to discover tens of thousands of local, resolved galaxies, greatly enhancing our knowledge of faint, nearby galaxy populations.

\begin{figure*}
\begin{center}
	\includegraphics[width=2.0\columnwidth]{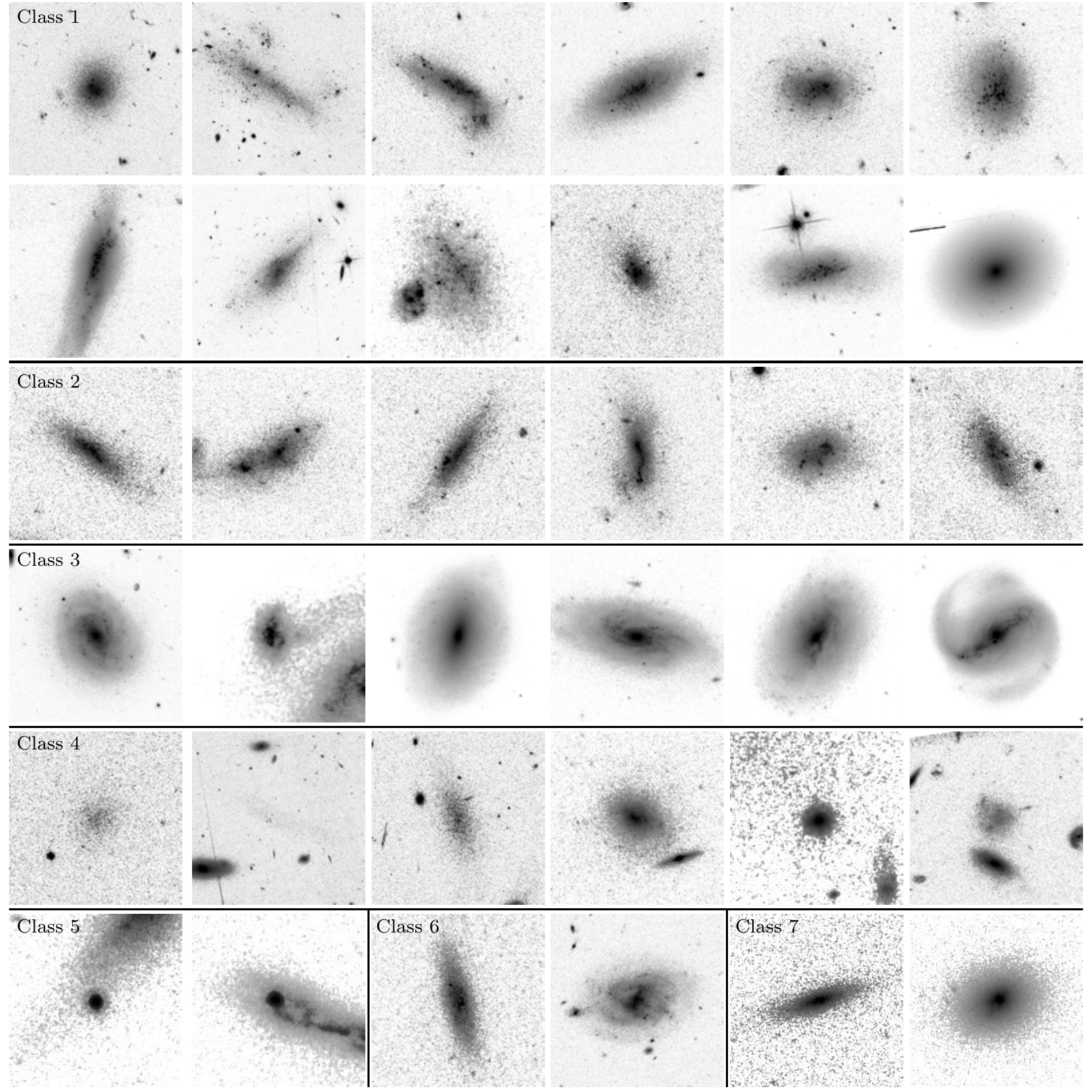}
	\caption{Cutouts {    from the COSMOS ACS mosaics \citep{Koekemoer2007}} showing examples of the different visual classes from the serendipitous sample. Each image is centred on the catalog coordinates and  scaled to 6.6$\,r_{\rm eff}$ on a side (with the exception of 709026, where the image is 15${\arcsec}$ on a side). Rows 1 and 2, from left to right, contain Class 1 (resolved) objects 213165, 260583, 331749, 401988, 458976, 561851, and 653748, 677414, 686606, 709026, 733922, 551648 respectively. (The last object on row 2, 551648 (ARK227), is Class 1 but may have the wrong spectroscopic distance.) Row 3 contains the Class 2 (marginally resolved) objects  259971, 279307, 589205, 627637, 642238, 997756. Row 4 contains the Class 3 (distant) objects 460674, 660791, 706494, 915194, 923647, 955856. Row 5 contains the Class 4 (LSB) objects 261496, 282078, 643833, 733610, 771819, 1038253. The final row contains Class 5 (contaminated) objects 377112, 484608, Class 6 (contaminated/LSB) objects 423926, 840592, and Class 7 (unclear) objects 518816, 731241.}
	\label{fig:cutouts}
	\end{center}
\end{figure*}

\begin{figure}
	 \includegraphics[width=\columnwidth]{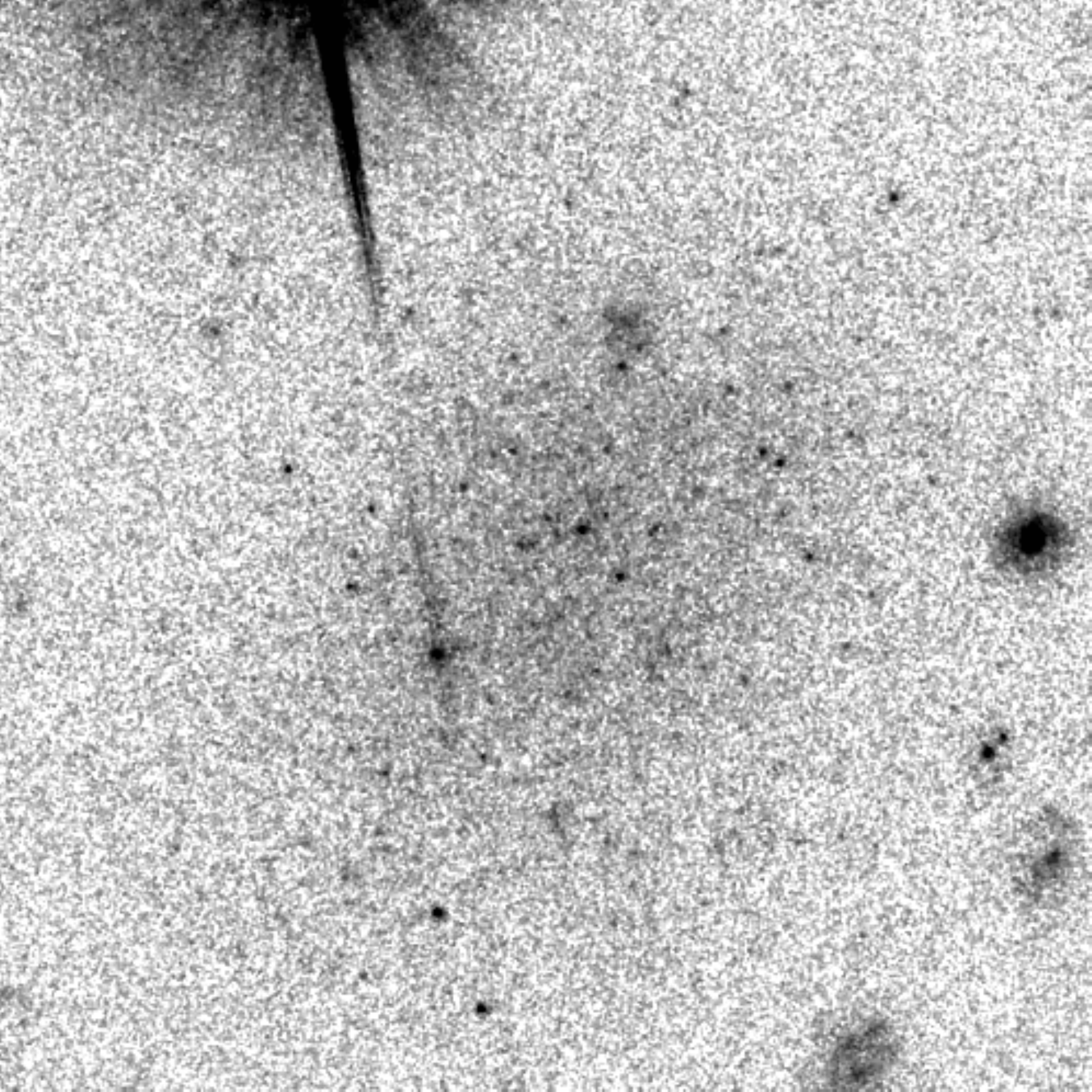}
    	\caption{An enlarged ACS F814W {    mosaic} image \citep{Koekemoer2007} 
	of one of the Class 1 objects (ID 549719 in the COSMOS 2015 catalogue), showing diffuse, low surface-brightness emission 
	and multiple point sources. The image is 15${\arcsec}$ on a side.}
        	\label{fig:DG1}
\end{figure}

\subsection{Notes on Individual Objects}

Table \ref{tbl:serendip} lists the IDs, coordinates, redshifts and magnitudes of the serendipitous discoveries, sorted by class. The IDs are from the COSMOS 2015 catalog, except where indicated. We note the following about individual objects:

\phantom{.}

{\noindent{\bf    260583} (LSBC L1-099) This is a bright Magellanic-type irregular, first catalogued by \cite{Impey96}, and detected in \mbox{H\,{\sc i}} by \cite{Taylor96}. It has a spectroscopic redshift of 1816 km/s, and is part of a dwarf-dominated group in the COSMOS field at a distance of roughly 26 Mpc. This galaxy is highly fragmented in the COSMOS 2015 catalogue; as many as 18 separate catalog entries may correspond to star-forming regions or nebulosity associated with this galaxy. }

\phantom{.}

{\noindent{\bf    279307} This irregular galaxy may be a superposition or merger between two or more objects. In the COSMOS 2015 catalogue, it is split into two separate components. It appears to have multiple faint/marginal point sources, so we have placed it in Class 2 (marginally resolved), although there is also a single, 
much brighter point source towards the edge of the object that could be a contaminating foreground star.}

\phantom{.}

{\noindent{\bf    549719} This low surface-brightness object, close to a bright star, is resolved into several dozen faint point sources in HST images (see Figure~\ref{fig:DG1}). Unusually,  it has imaging in three separate ACS filters, F814W and F475W from the COSMOS survey, and F606W (as well as F814W) from the CANDELS survey \citep{CANDELS}.
A comparison of the different HST images shows that the point sources have a broad range of colours, suggesting that they may be the brightest (supergiant) stars in an actively star-forming system. The object also appears bright in GALEX images of the COSMOS field.
The photometric redshift puts this object at a rough distance of 21.5\,$\pm$\,34 Mpc, but given many {    COSMOS galaxies} in this distance range are part of the previously mentioned group at 26 Mpc, it seems likely that this object is another faint member of the group. If one or the other of these two distance estimates is correct, 549719 has an absolute magnitude of $-12.4$ or $-12.7$ in $i^{+}$, making it one of the faintest resolved galaxies known at this distance.

On the other hand, another intriguing possibility is that 549719 could be a nearby analogue of the `ultra-diffuse galaxies' (UDGs) recently discovered in the Coma cluster \citep{vanDokkum2015}. Slightly deeper {\it HST} imaging of these objects shows them to be diffuse, low surface-brightness, roughly spheroidal systems, with dozens of bright point sources corresponding to globular clusters \citep{vanDokkum2017}. While the point sources in 549719 show a broad range of colours, and depending on its distance, may be too faint to be globular clusters, the possibility that this object is a field UDG warrants further investigation, as it does for several other objects in the serendipitous catalogue (e.g.~458976, 316142,  the very faint 300323, and the objects in Class 4).}

\phantom{.}

{\noindent{\bf    551648} (ARK227) This previously catalogued galaxy has a spectroscopic redshift of 1793 km/s,  putting it at a distance of $D \sim$26\,Mpc, in the same group as 26058 and 677414, and giving it an absolute magnitude of $-17.8$. It seems likely this redshift is incorrect, however, as the galaxy appears to be an intrinsically bright, regular elliptical with a large population of globular clusters. The brightest of these have magnitudes of $i^{+}\,\sim\,$23--24, suggesting a distance up to two times further away. }

\phantom{.}

{\noindent{\bf    677414} (LSBC L1-100) This is another bright, Magellanic-type irregular, originally catalogued by \cite{Impey96}. It has a spectroscopic redshift of 1729 km/s, and is likely part of the same group as 260583. It is fragmented into at least four separate components in the COSMOS 2015 catalogue.}

\phantom{.}

{\noindent{\bf    709026} The size of this object appears to be incorrect in the COSMOS 2015 catalogue, so we have included a 15${\arcsec}$ cutout in Figure~\ref{fig:cutouts}. It has many distinct point sources, however, as well as extended diffuse emission, so it is clearly Class 1.}  

\phantom{.}

{\noindent{\bf    J100222.70+022520.3} This object is large and relatively diffuse, but is also located very close to a bright star. {    In the deeper ground-based COSMOS images it appears to have a central bar and twisting isophotes.} It was masked out of the COSMOS 2015 catalogue, although it appears in earlier versions of the COSMOS {    photometric and photo-$z$ catalogues \citep[][]{Capak2007, Mobasher07}, where it has a photo-$z$ of 0.09 (i.e.~a distance of $D \sim$400\,Mpc)}. It is not clear whether it contains resolved point sources; the one or two in this area may be foreground stars seen in projection. Given its unusual size and surface brightness, we have included it in Class 7, {    although it is also another plausible candidate field UDG}.}

\phantom{.}

Finally, we note that two objects, 213165 and 259971, have multiple conflicting redshifts listed within 1${\arcsec}$ of each other. 213165 has redshifts 0.03 and 0.1529 listed, while 
259971 has redshifts 0.01 and 0.8058 listed. Both objects appear to be local, however (particularly 213165, which appears to be resolved into multiple point sources), so the status of these objects is unclear. 

\section{Summary and Conclusions}
\label{sec:8}

From a theoretical point of view, dwarf galaxies are particularly important as tracers of small-scale dark matter structure, both in the field and within the halos of brighter galaxies. The faintest identified dwarfs are members of the Local Group, but this sample may not be representative of dwarf properties in general. In particular, if satellite populations depend on the properties -- stellar mass, morphology, and/or detailed star formation history -- of their central galaxy, many more examples of satellite populations 
will be needed to clarify these connections. Thus, identifying intrinsically faint galaxies in the nearby universe beyond the Local Group is of considerable importance. 

Based on the local samples that exist, there should be a large population of objects just below the spectroscopic limits of current wide-field surveys, whose distinct {    structural properties} can be used to separate them to some degree from the much larger number of faint background galaxies. In this paper, we have experimented with {    structural} selection as a tool for quantifying local dwarf populations, selecting samples with cuts in magnitude, size and surface brightness, and using their clustering with respect to bright galaxies with known redshifts to confirm that some fraction of the selected sample is indeed nearby. 

We have tested this approach using the photometric redshift catalog of the COSMOS survey, since it is one of the only sources of accurate redshift estimates for large numbers of faint galaxies. In other ways, however, COSMOS is not the ideal survey for our purposes, as it covers only a small field. As a result, we have pushed our approach, originally introduced in \citetalias{ST14} to identify galaxies with $\sim$40 Mpc, out to a redshift of $z=0.15$ or more, that is roughly 15 times further away. 

We find that {    structural} selection does work surprisingly well even out to these distances, although it starts to fail beyond that. It produces samples enhanced in local dwarfs that are neither complete nor unbiased in magnitude or luminosity, but can nonetheless be useful in studying satellite abundance at a statistical level. Our best selection cuts recover two-thirds of the clustering signal measured using the extremely high quality COSMOS photo-$z$s, with 80\%\ of the SNR, and a purity of $\sim$33\%.

The {    structural} selection methods tested here were deliberately based on SDSS photometry in the COSMOS field, which has both poor spatial resolution and fairly bright surface-brightness limits ($\left<\mu\right>_{\rm eff} \lesssim$ 25--27). 
A new generation of low surface-brightness instruments (e.g.~The Dragonfly Telephoto Array -- \citealt{Dragonfly}) and/or surveys 
(e.g.~MATLAS\footnote{http://irfu.cea.fr/Projets/matlas/MATLAS/MATLAS.html}, 
LSST\footnote{https://www.lsst.org}, HSC--SSP \citep{HSC})
will push detection limits much further into the region of parameter space populated by the known local dwarfs. {    \cite{Danieli2018}, for instance,  
show that integrated light surveys with Dragonfly or similar instruments could detect typical local dwarfs in the magnitude range $M_V = -5$ to $-10$ out to distances of $D \sim10$\,Mpc. Thus, these surveys will fill in the gap between the `ultrafaints', detected locally using star counts, and the brighter populations we are able to characterize at larger distances ($z \le 0.15$, or $D \lesssim 600$\,Mpc), using {    structural} selection and clustering. We note, however, that spectroscopic follow-up may be challenging or impossible for very low surface-brightness objects, so even with these new samples, clustering analysis may still be required to determine the purity and true satellite fraction.}

Our serendipitous discovery of dozens of local galaxies in the COSMOS field also augurs well for future space-based imaging surveys. While the COSMOS samples of very local galaxies are relatively small, surveys such as 
Euclid\footnote{http://sci.esa.int/euclid} or WFIRST\footnote{https://wfirst.gsfc.nasa.gov} should  detect tens of thousands of similar objects. Here too, we expect {    structural} selection to help significantly in separating nearby galaxies from distant ones, revealing the faintest galaxies of the local universe.

\section*{Acknowledgements}

The authors acknowledge useful discussions with Simon Driver (on methods to identify nearby galaxies) and Alexandar Mechev (on the likely distance to ARK227). {    We also thank the referee, Bob Abraham, for a number of helpful comments and for pointing out the possible connection to UDGs.} Finally, we thank our friends and collaborators from the COSMOS survey for many years of support and advice, and for first pointing out many of the objects in the serendipitous catalogue.

This paper made use of the NASA Extragalactic Database (NED -- \url{http://ned.ipac.caltech.edu}), the COSMOS cutout service at IRSA (\url{http://irsa.ipac.caltech.edu/data/COSMOS}), Knud Jahnke's COSMOS Skywalker visual search engine (\url{https://www.mpia.de/COSMOS/skywalker}), {    and Stephen Gwyn's interface to the multi-wavelength coverage in the COSMOS field (\url{http://www.cadc-ccda.hia-iha.nrc-cnrc.gc.ca/en/megapipe/cfhtls/scrollD2.html})}. We thank the creators of these resources for facilitating this work.

JET acknowledges support from the Natural Science and Engineering Research Council of Canada, through a Discovery Grant. JR was supported by JPL, which is run under a contract for NASA by Caltech. RM is supported by a Royal Society University Research Fellowship.

The COSMOS 2015 catalog is based on data products from observations made with ESO Telescopes at the La Silla Paranal Observatory under ESO programme ID 179.A-2005 and on data products produced by TERAPIX and the Cambridge Astronomy Survey Unit on behalf of the UltraVISTA consortium.

\bibliographystyle{mnras}
\bibliography{morpho}

\appendix
\section{The Serendipitous Catalogue}

In Table~\ref{tbl:serendip} we list the serendipitous catalogue of nearby objects. Columns are visual class (as explained in section~\ref{sec:7}), ID from the COSMOS 2015 catalogue (where available), coordinates, redshift, redshift error (for objects with photometric redshifts only), apparent $i^+$-band magnitude, approximate absolute magnitude in the same band (assuming a distance $D = cz/H_0$ with $H_0 = 0.678$), and any comments. As noted previously, repeated visual searches suggest classes 1 \& 2 are reasonably complete, while classes 3--7 contain only a few representative examples of the many objects of this kind.

\begin{table*}
	\centering		
	\caption{The Serendipitous Catalog} 
	\begin{tabular}{rcccllccl}
		\hline\hline
		Class & COSMOS ID & R.A. & Decl. & $z$ & $\sigma_z^*$ & $i^+$ & $M_{i^+}$ &Comments\\
		& (\citealt{Laigle16})  & $(J2000)$ & (J2000) & & & (mag) & (mag) &\\
		\hline
1 & 213165 & 150.6950 & 1.6139 & 0.030 &  & 18.02 & -17.5 & conflicting redshift 0.1529\\
 & 260583 & 149.6202 & 1.6936 & 0.006 &  & 17.90 & -14.2 & part of group at 26 Mpc\\
 & 331749 & 150.3456 & 1.7936 & 0.019 &  & 18.96 & -15.6 \\
 & 401988 & 150.0245 & 1.9110 & 0.006 &  & 17.15 & -15.0 & part of group at 26 Mpc\\
 & 458976 & 149.8663 & 2.0071 & 0.013 &  & 18.59 & -15.2 \\
 & 549719 & 150.1254 & 2.1498 & 0.005 & 0.008 & 19.00 & -12.7 \\
 & 551648 & 150.0433 & 2.1560 & 0.006 &  & 14.24 & -17.8 & appears more distant?\\
 & 561851 & 150.6131 & 2.1668 & 0.006 &  & 18.02 & -14.1 & part of group at 26 Mpc\\
 & 653748 & 150.3134 & 2.3064 & 0.027 &  & 17.42 & -17.9 & \\%
 & 677414 & 149.6951 & 2.3477 & 0.006 &  & 17.37 & -14.6 & part of group at 26 Mpc\\\
 & 686606 & 150.3666 & 2.3404 & 0.007 &  & 20.28 & -12.2 & part of group at 26 Mpc\\\
 & 709026 & 150.0284 & 2.3793 & 0.012 &  & 19.45 & -14.1 & size incorrect in catalogue?\\
 & 733922 & 150.4743 & 2.4138 & 0.007 &  & 17.40 & -14.9 & part of group at 26 Mpc \\
2 & 219550 & 149.8758 & 1.6103 & 0.040 & 0.034 & 19.87 & -16.3 \\
 & 221686 & 149.5820 & 1.6156 & 0.043 & 0.035 & 21.77 & -14.6 \\
 & 259971 & 149.4614 & 1.6750 & 0.010 &  & 20.20 & -13.0 & conflicting redshift of 0.8058\\
 & 279307 & 149.9644 & 1.7067 & 0.025 &  & 19.60 & -15.6 \\
 & 300323 & 150.4282 & 1.7425 & 0.045 & 0.045 & 22.28 & -14.1 \\
 & 316142 & 149.4853 & 1.7645 & 0.018 & 0.029 & 20.19 & -14.2 \\
 & 424575 & 149.5127 & 1.9533 & 0.005 & 0.008 & 17.97 & -13.7 \\
 & 556961 & 149.6577 & 2.1597 & 0.005 & 0.008 & 20.84 & -10.8 \\
 & 589205 & 149.8118 & 2.1923 & 0.025 &  & 19.92 & -15.2 \\
 & 627637 & 149.7679 & 2.2548 & 0.025 &  & 19.07 & -16.1 \\
 & 642238 & 149.4566 & 2.2722 & 0.005 & 0.008 & 19.93 & -11.8 \\
 & 689831 & 150.6784 & 2.3433 & 0.005 & 0.008 & 21.32 & -10.3 \\
 & 880363 & 149.9964 & 2.6334 & 0.060 & 0.040 & 21.10 & -16.0 \\
 & 918161 & 150.3921 & 2.6917 & 0.012 &  & 20.25 & -13.3 \\
 & 989145 & 150.4089 & 2.8052 & 0.044 &  & 20.26 & -16.1 \\%
 & 997756 & 149.6831 & 2.8163 & 0.023 & 0.023 & 20.19 & -14.8 \\
3 & 183741 & 149.5938 & 1.5848 & 0.028 &  & 15.89 & -19.5 \\
 & 246757 & 149.4982 & 1.6542 & 0.022 &  & 19.83 & -15.0 \\
 & 460674 & 150.5469 & 2.0216 & 0.021 &  & 16.21 & -18.6 \\
 & 532836 & 150.5065 & 2.1134 & 0.046 &  & 18.78 & -17.7 \\
 & 534651 & 150.1830 & 2.1148 & 0.100 & 0.060 & 21.62 & -16.5 \\
 & 538389 & 150.0464 & 2.1188 & 0.029 &  & 21.86 & -13.6 \\
 & 622498 & 150.1930 & 2.2445 & 0.677 &  & 21.30 & -21.0 \\
 & 660791 & 149.9128 & 2.3040 & 0.705 &  & 21.91 & -20.5 \\
 & 706494 & 150.2301 & 2.3955 & 0.045 &  & 15.44 & -21.0 \\
 & 718332 & 149.8389 & 2.3875 & 0.028 &  & 22.17 & -13.2 \\
 & 824852 & 149.7570 & 2.5499 & 0.029 &  & 18.66 & -16.8 \\
 & 905622 & 150.4302 & 2.6859 & 0.047 &  & 17.31 & -19.2 \\
 & 915194 & 149.8467 & 2.6938 & 0.048 &  & 16.78 & -19.8 \\
 & 923647 & 150.0386 & 2.7132 & 0.033 &  & 16.77 & -19.0 \\
 & 955856 & 150.0338 & 2.7651 & 0.029 &  & 15.41 & -20.1 \\
4 & 261496 & 149.5315 & 1.6786 & 0.021 & 0.026 & 20.63 & -14.1 \\
 & 282078 & 149.8230 & 1.7285 & 0.055 & 0.183 & 20.18 & -16.7 \\
 & 643833 & 149.9028 & 2.2784 & 0.005 & 0.008 & 19.86 & -11.8 \\
 & 733610 & 150.1712 & 2.4130 & 0.043 &  & 19.50 & -16.9 \\%
 & 771819 & 150.3126 & 2.4689 & 0.005 & 0.008 & 20.92 & -10.7 \\
 & 1038253 & 149.8371 & 2.8744 & 0.050 & 0.035 & 20.08 & -16.6 \\
5 & 377112 & 150.1917 & 1.8634 & 0.027 &  & 20.84 & -14.5 \\
 & 484608 & 150.4819 & 2.0372 & 0.005 & 0.008 & 20.66 & -11.0 \\
 & 494700 & 150.4874 & 2.0533 & 0.093 &  & 20.86 & -17.1 \\
 & 648571 & 150.3759 & 2.2856 & 0.051 & 0.036 & 20.76 & -15.9 \\
 & 864285 & 150.6092 & 2.6075 & 0.104 &  & 20.68 & -17.6 \\
		\hline
		&\multicolumn{6}{c}{$^*$ redshift error, listed only for objects with photometric redshifts}\\
	\end{tabular}
	\label{tbl:serendip}
\end{table*}

\begin{table*}
	\centering		
	\contcaption{The Serendipitous Catalog} 
	\begin{tabular}{rcccllccl}
		\hline\hline
		Class & COSMOS ID & R.A. & Decl. & $z$ & $\sigma_z^*$ & $i^+$ & $M_{i^+}$ &Comments\\
		& (\citealt{Laigle16})  & $(J2000)$ & (J2000) & & & (mag) & (mag) &\\
		\hline
6 & 380820 & 150.0600 & 1.8665 & 0.024 &  & 20.04 & -15.0 \\
 & 423926 & 150.3431 & 1.9400 & 0.046 &  & 20.08 & -16.4 \\%
 & 532809 & 150.7758 & 2.1105 & 0.005 & 0.008 & 18.45 & -13.3 \\
 & 840592 & 150.7351 & 2.5780 & 0.005 & 0.008 & 17.78 & -14.0 \\
 & 880547 & 150.0023 & 2.6332 & 0.024 &  & 20.20 & -14.9 \\
7 & 216843 & 149.6873 & 1.6104 & 0.050 & 0.035 & 19.93 & -16.7 \\
 & 349181 & 149.8123 & 1.8196 & 0.081 & 0.051 & 19.39 & -18.3 \\
 & 400833 & 150.7306 & 1.9004 & 0.005 & 0.008 & 21.05 & -10.7 \\
 & 516283 & 150.6366 & 2.0837 & 0.093 &  & 20.92 & -17.1 \\
 & 518816 & 150.7234 & 2.0883 & 0.069 &  & 21.37 & -16.0 \\%
 & 523477 & 150.4045 & 2.1067 & 0.054 & 0.040 & 18.10 & -18.7 \\
 & 731241 & 150.1731 & 2.4042 & 0.036 & 0.032 & 19.32 & -16.6 \\
 & 757311 & 150.0542 & 2.4513 & 0.082 &  & 20.38 & -17.4 \\%
 & 837992 & 150.6170 & 2.5750 & 0.007 & 0.024 & 18.84 & -13.5 \\
 & 840823 & 150.4003 & 2.5727 & 0.092 &  & 21.09 & -16.9 \\%
 & 862172 & 149.7740 & 2.6061 & 0.188 &  & 21.47 & -18.0 \\%
 & 908277 & 150.7575 & 2.6765 & 0.005 & 0.008 & 21.56 & -10.1 \\
 & 943231 & 150.0981 & 2.7438 & 0.059 & 0.040 & 19.21 & -17.8 \\
 & (masked) & 150.5946 & 2.4223 & 0.090 & 0.090 & 21.46$^\dagger$& -16.5 & SDSS J100222.70+022520.3\\
  &&&&&&&& {    $^\dagger$SDSS $i$-band {\it model} magnitude}\\
\\
		\hline
		&\multicolumn{6}{c}{$^*$ redshift error, listed only for objects with photometric redshifts}\\
	\end{tabular}
	\label{tbl:serendipcont}
\end{table*}

\bsp	
\label{lastpage}
\end{document}